%% file: TW-Dec-11-2304.tex
\documentclass[journal,twoside,final]{IEEEtran}

\usepackage{comment}
\usepackage{url}

\usepackage{dsfont}
\usepackage{amssymb}
\usepackage{stfloats}
\usepackage{graphicx}
\usepackage[cmex10]{amsmath}
\usepackage{array}
\usepackage{epsfig}
\usepackage[table]{xcolor}
\usepackage[noadjust]{cite}
\usepackage{hyperref}
\hypersetup{
    colorlinks,%
    citecolor=black,%
    filecolor=black,%
    linkcolor=black,%
    urlcolor=black
}

\usepackage{flushend}
\begin{document}
\newcommand {\SINR} {\textrm{SINR}}
\title{Unified Performance Analysis of Orthogonal Transmit Beamforming Methods with User Selection}
\author{Serdar~Ozyurt
        and Murat~Torlak,~\IEEEmembership{Senior Member,~IEEE}\\
\thanks{%
Manuscript received January 11, 2012; revised September 14, 2012; accepted October 30, 2012. The associate
editor coordinating the review of this paper and approving it for
publication was X. Dong.}
\thanks{%
This paper was presented in part at IEEE Global Communications Conference, 2011, Houston, TX, USA.}
\thanks{%
S. Ozyurt was with the Department of Electrical Engineering, University of Texas at Dallas, Richardson, TX, USA. He is now with the Department of Energy Systems Engineering, Yildirim Beyazit University, Ankara, Turkey (e-mail: sozyurt@ybu.edu.tr).}
\thanks{%
M. Torlak is with the Department of Electrical Engineering, University of Texas at Dallas, Richardson, TX, USA (e-mail: torlak@utdallas.edu).}
\thanks{%
Digital Object Identifier 10.1109/TWC.2012.112304.}}

\maketitle

\markboth{IEEE Transactions on Wireless Communications, Vol. XX, No.
XX, Month 2012}{Ozyurt \MakeLowercase{and} Torlak:
 Unified Performance Analysis of Orthogonal Transmit Beamforming Methods with User Selection}

\pubid{1536-1276/12\$31.00~\copyright~2012 IEEE}

\pubidadjcol
\begin{abstract}
Simultaneous multiuser beamforming in multiantenna downlink channels can entail dirty paper (DP) precoding (optimal and high complexity) or linear precoding (suboptimal and low complexity) approaches. The system performance is typically characterized by the sum capacity with homogenous users with perfect channel state information at the transmitter. The sum capacity performance analysis requires the exact probability distributions of the user signal-to-noise ratios (SNRs) or signal-to-interference plus noise ratios (SINRs). The standard techniques from order statistics can be sufficient to obtain the probability distributions of SNRs for DP precoding due to the removal of known interference at the transmitter. Derivation of such probability distributions for linear precoding techniques on the other hand is much more challenging. For example, orthogonal beamforming techniques do not completely cancel the interference at the user locations, thereby requiring the analysis with SINRs. In this paper, we derive the joint probability distributions of the user SINRs for two orthogonal beamforming methods combined with user scheduling: adaptive orthogonal beamforming and orthogonal linear beamforming. We obtain compact and unified solutions for the joint probability distributions of the scheduled users' SINRs. Our analytical results can be applied for similar algorithms and are verified by computer simulations.
\end{abstract}


\section{Introduction}
\PARstart{I}{n} a downlink multiuser multiple-input single-output (MISO) system where a base station (BS) equipped with $M$ antennas communicates with $K$ single-antenna users, theoretical results demonstrate that the sum capacity scales linearly with $\min\{K, M\}$. In particular, if perfect channel state information (CSI) is available at BS, it has been shown that a method based on dirty paper coding (DPC) achieves the channel capacity~\cite{Weingarten:06}. The implementation of the DPC technique is a challenging task for the present as it requires complicated coding operations and an exhaustive search to schedule an optimal set of users when $K>M$. Alternative linear precoding techniques in the form of multiuser beamforming (BF) at BS (as known as space-division multiple access) can be employed to serve users separately where each scheduled user's data stream is individually encoded and multiplied by a special BF weight vector. Despite independent encoding among users, the computation of the optimal BF weight vectors and user scheduling can still be problematic. A number of suboptimal BF methods combined with user scheduling have been proposed to handle user scheduling and BF weight vector design problems jointly at relatively low complexity~\cite{Yin:02}-\cite{Duplicy:07}. The suboptimal schemes in~\cite{Yoo:06}-\cite{Duplicy:07} have been shown to achieve a significant portion of the DPC sum capacity. A joint BF and user scheduling scheme (OLBF) with a complexity of ${O}(M^{2}K)$ has been described in~\cite{Francisco:07} and shown to outperform the zero-forcing based scheme of~\cite{Dimic:05} in the sum rate sense at low signal-to-noise ratio (SNR) values for low to moderate number of users. A similar scheme has been reported in~\cite{Duplicy:07} with a higher complexity of ${O}(M^{3}K)$ where the joint determination of BF weight vectors and user scheduling is carried out in a sequential manner. This scheme (adaptive OBF) has been shown to yield an increased sum rate performance as compared to the scheme of~\cite{Francisco:07}. An important advantage of using orthogonal BF weight vectors is that it allows the scheduler to easily express and compute exact signal-to-interference plus noise ratio (SINR) values during user scheduling process~\cite{Duplicy:07}.\pubidadjcol

The standard techniques from order statistics can be sufficient to obtain probability distributions of SNR values and the sum capacity performance for DPC-based precoding schemes due to the removal of known interference and use of suboptimal but effective user selection algorithms~\cite{Dimic:05}. Most of the performance analysis studies on more practical multiuser BF algorithms are on the other hand based upon asymptotic analysis. An exact statistical analysis is essential to gain a complete insight into the system performance and to analytically determine some useful parameters~\cite{Ozdemir:10,Ozyurt&VBLAST:11,Ozyurt:12}. The difficulty with the performance analysis of linear precoding schemes with user scheduling mainly arises from the fact that user ordering changes the statistics of the users making it quite complicated to express mathematically. Also, depending on the selection procedure, a decision at any point during user scheduling can deeply alter the statistics of the previously scheduled users~\cite{Ozyurt:12}. A multiuser BF scheme with a greedy user selection algorithm has been analyzed in~\cite{Dimic:05}. This scheme combines scalar DPC with linear precoding to completely suppress the interference at the transmitter. Using conditional probability density function (PDF) expressions, the authors provide a compact solution for the marginal PDFs of the scheduled users' channel gains. Note that even in this no-interference case, a direct approach with conditional distributions requires complicated algebraic operations to find a general formula. In~\cite{Huang:09}, a joint user scheduling and BF algorithm relied on orthogonal BF and limited feedback is analyzed. The asymptotic sum rate scaling laws are obtained assuming an asymptotically large number of users. Another low-complexity suboptimal scheme where orthogonal BF weight vectors are formed in a random manner is analyzed in~\cite{Yang:11}. The authors present exact results for the PDFs of the selected users' SINRs.

In this paper, we develop statistical tools for analyzing the sum rate performance of two multiuser orthogonal BF algorithms by assuming single-antenna users and perfect CSI at the transmitter. By approaching the problem from a different angle than the previous approaches, we derive an exact and general result from order statistics on the joint PDF of scheduled users' SINRs for a given orthogonal BF algorithm combined with a greedy user scheduling technique. Our approach is built upon the joint PDF of unordered SINRs with random user selection where the orthogonality of BF weight vectors facilitates the analysis. The difference from the work in~\cite{Yang:11} is that our result is given for general greedy user scheduling algorithms under certain conditions. We apply this result specifically to adaptive OBF and OLBF to find compact and unified solutions for the joint PDF of the scheduled users' SINRs. We verify our analysis with numerical results.

Notation: We use uppercase and lowercase boldface letters to represent matrices and vectors, respectively. The operators $\mathds{E}[.]$, $(.)^{H}$, $|.|$, $\|.\|$, $\text{Pr}(.)$, $\setminus$, and $\cup$ denote expectation, Hermitian transpose, cardinality (absolute value for scalars), Euclidean norm, probability, set difference, and set union, respectively. Also, $\textbf{A}(:,n)$ and $\textbf{A}(:,1:n)$ represent a vector equal to $n$th column of the matrix $\textbf{A}$ and a matrix comprised of first $n$ columns of the matrix $\textbf{A}$, respectively. We refer to the joint PDF of $\{\alpha_1, \ldots, \alpha_n \}$ by $f(\alpha_1, \ldots, \alpha_n)$ with $f(\alpha_1 = a_1, \ldots, \alpha_n = a_n)$ representing the joint PDF of $\{\alpha_1, \ldots, \alpha_n \}$ evaluated at $\{a_1, \ldots, a_n \}$. Additionally, $F_{\beta_n}(\beta_1, \ldots, \beta_n)$ denotes the joint cumulative distribution function (CDF) of $\{\beta_1, \ldots, \beta_n\}$ and $F_{\beta_n}(b_1, \ldots, b_n)$ is the joint CDF of $\{\beta_1, \ldots, \beta_n\}$ evaluated at $\{b_1, \ldots, b_n\}$.

The rest of the paper is organized as follows: Section~\ref{sytemmodel} presents the system model and a key result for statistical analysis of the joint orthogonal BF and user scheduling algorithms under certain conditions is derived from order statistics in Section~\ref{mainresult}. As applications of this result, Sections~\ref{adaptiveOBF} and~\ref{OLBF} are dedicated to the statistical analysis of adaptive OBF and OLBF, respectively. Numerical results are presented in Section~\ref{numericalresults} and Section~\ref{conclusion} concludes the paper.
\section{System Model}
\label{sytemmodel}
A single cell MISO broadcast channel with $M$ antennas at BS and $K$ ($K\geq M$) single-antenna users is considered where the total available transmit power is given by $P$. Assuming perfect CSI and equal power allocation among the scheduled users at BS, the received signal at $i$th scheduled user $d_{k_{i}}$ can be expressed as follows
\begin{equation}
d_{k_{i}}=\sqrt{\frac{P}{\left|{U_{n}}\right|}}\textbf{h}_{k_{i}}^{H}
\textbf{w}_{k_{i}}l_{k_{i}}+\sqrt{\frac{P}{\left|{U_{n}}\right|}}\textbf{h}_{k_{i}}^{H}
\sum_{j \in U_{n}, j\neq i}\textbf{w}_{k_{j}}l_{k_{j}}+e_{k_{i}}
\label{eq:kanalmodeli}
\end{equation}
where $\textbf{h}_{k_{i}}^{H}=\left[h_{k_{i}}(1)~h_{k_{i}}(2)~\ldots~h_{k_{i}}(M)\right]$ is the channel vector, $\textbf{w}_{k_{i}} \in \mathbb{C}^{M\times 1}$ is the unit-norm beamforming weight vector, $l_{k_{i}}\in \mathbb{C}$ is the unit-norm transmitted data symbol, and $e_{k_{i}}\in \mathbb{C}$ is additive white Gaussian noise with zero mean and unit variance for $i$th scheduled user, respectively. In (\ref{eq:kanalmodeli}), the set of scheduled users is denoted by $U_{n}$ where the number of scheduled users is given by $\left|{U_{n}}\right|=n$. We assume slow flat fading and a homogeneous network where the elements of $\textbf{h}_k$ ($k$th user's channel vector with $k \in \{1,2,\ldots,K\}$) are independent and identically distributed (IID) zero-mean complex Gaussian random variables with unit variance.
\section{\!A New Look at Order Statistics for Statistical Analysis of Joint Greedy User Scheduling and Orthogonal Beamforming Algorithms}
\label{mainresult}
In this section, we obtain a key result from order statistics on the joint PDF of scheduled users' SINRs for a given greedy user scheduling and orthogonal BF algorithm. We assume that $r$ users with $\{K,M\} \geq r \geq 2$ are sequentially scheduled based on their SINRs, i.e., user with the maximum SINR is scheduled at each step. The scheduled users' SINRs are denoted by $\{y_1, y_2, \ldots, y_r\}$. In order to find the joint PDF of $\{y_1, y_2, \ldots, y_n\}$ for\linebreak $n \in \{1,2,\ldots,r\}$, we first form the joint PDF of unordered SINRs temporarily ignoring ordering among users due to user scheduling where we assume each candidate user has the same likelihood to be selected at each step (random user selection). This is given by $f(v_{k1},v_{k2},\ldots,v_{kn})$ for a certain set of rules (or region) denoted by $S_k$ for $k$th user and can be obtained under a certain unitary transformation. The variable $v_{kj}$ represents resulting SINR when $k$th user $(k \in \{1, 2,\ldots, K\})$ is scheduled at $j$th step\linebreak $(j \in \{1, 2,\ldots, n\})$ when no ordering is applied at any step. Also, $\{v_{k1},v_{k2},\ldots,v_{kn}\}$ represent the candidacy SINR of $k$th user at the steps $\{1,2,\ldots,n\}$, respectively. The set of rules $S_k$ depends on the algorithm, e.g. $v_{k1} \geq v_{k2} \geq \ldots \geq v_{kn} \geq 0$. We can write the following composite joint PDF including all $K$ users as
\begin{flalign}
f\left(\{v_{k1},v_{k2},\ldots,v_{kn}\}_{k=1}^{K}\right) && \nonumber
\end{flalign}
\begin{equation}
=\left\{ \begin{array}{lc}
\prod_{k=1}^{K} f(v_{k1},v_{k2},\ldots,v_{kn}) & \mbox{\!\!for~~} S_1, S_2, \ldots, S_K,\\
0 & \mbox{otherwise},
       \end{array} \right.
       \label{eq:compositejointpdf}
\end{equation}
where independence among $K$ joint PDFs stems from no ordering assumption. Now, we take the user ordering (due to user scheduling) into account by initially assuming that $v_{11}$ is scheduled at the first step, i.e., $y_{1}=v_{11}=\underset{k}{\max}~v_{k1}$. It is worth to mention that $v_{11}$ and $\{v_{k1}\}_{k=2}^{K}$ are no longer independent. In a similar way, we assume that $v_{22}$ is scheduled at the second step, i.e., $y_{2}=v_{22}=\underset{k\geq 2}{\max}~v_{k2}$. Note that due to the first scheduling step, the variables with the subscript $k=1$ are disregarded for the subsequent steps. Repeating this scheduling process for $n$ steps and applying Bapat-Beg theorem from order statistics~\cite{David:03}, the new composite joint PDF can be written as
\begin{flalign}
f\left(\{v_{k1},v_{k2},\ldots,v_{kn}\}_{k=1}^{K}|_{\substack{v_{11}=y_1,\\v_{22}=y_2, \ldots, v_{nn}=y_n}}\right) && \nonumber
\end{flalign}
\begin{equation}
=\left\{ \begin{array}{l}
\frac{K!}{(K-n)!} \prod_{k=1}^{K} f\left(\{v_{k1},v_{k2},\ldots,v_{kn}\}|_{\substack{v_{11}=y_1,\\v_{22}=y_2, \ldots, v_{nn}=y_n}}\right) \vspace{.05in} \\ \hspace{1in}\mbox{~~for~~} S_1, S_2, \ldots, S_K, \mbox{~and~} \cup_{i=1}^{n}R_i, \vspace{.1in} \\
 0 \hspace{2in}\mbox{otherwise},
\end{array} \right.
\label{eq:esit1}
\end{equation}
with
\begin{equation}
R_{i} : \left\{y_{i} \geq  \{v_{i+1,i}, v_{i+2,i},\ldots, v_{Ki}\}\right\}\text{~~for~~}i \in \{1,2,\ldots,n\}.
\label{eq:sinirlar}
\end{equation}
In (\ref{eq:esit1}), $K!/(K-n)!$ is the number of different $n-$permutations that can be selected out of $K$ users. Consequently, the joint PDF of $\{y_{1},y_{2},\ldots,y_{n}\}$ can be obtained from
\begin{flalign}
f(y_{1},y_{2},\ldots,y_{n}) && \nonumber
\end{flalign}
\vspace{-.15in}
\begin{equation}
=\!\!\int\limits_{S_C}\!\!f\left(\{v_{k1},v_{k2},\ldots,v_{kn}\}_{k=1}^{K}|_{\substack{v_{11}=y_1,\\v_{22}=y_2,\ldots, v_{nn}=y_n}}\right) \!\prod_{k=1}^{K} \prod_{\substack{
            j=1\\
j\neq k}}^{n} d v_{kj}
\label{eq:esit2}
\end{equation}
where the integration region $S_C$ is defined by union of $S_1, S_2, \ldots, S_K$, and $\cup_{i=1}^{n}R_i$. Substituting (\ref{eq:esit1}) and (\ref{eq:sinirlar}) into (\ref{eq:esit2}) and applying some manipulations, we can write
\begin{flalign}
f(y_{1},y_{2},\ldots,y_{n})=\frac{K!}{(K-n)!}\left(\prod_{k=1}^{n}\phi_k(y_k,
y_{k-1},\ldots,y_1)\right) && \nonumber
\end{flalign}
\begin{equation}
\hspace{1.1in} \times
\left(\prod_{k=n+1}^{K}F_{v_{kn}}(y_{1},y_{2},\ldots,y_{n})\right)
\label{eq:mainresult}
\end{equation}
where $\phi_1(y_1)=f(v_{11}=y_1)$ and
\begin{flalign}
\phi_k(y_k,
y_{k-1},\ldots,y_1)=\hspace{-.15in}\int\limits_{S_{k}\cup\left(\cup_{i=1}^{k-1} R_i\right)}\hspace{-.33in}f(v_{k1},v_{k2},\ldots,v_{k,k-1},v_{kk}=y_k)  && \nonumber
\end{flalign}
\vspace{-.25in}
\begin{flalign}
&& \times dv_{k1}dv_{k2}\ldots dv_{k,k-1} \nonumber
\end{flalign}
for $k \in \{2, 3, \ldots, n\}$. Also, $f(v_{k1},v_{k2},\ldots,v_{kk})$ is the joint PDF of $\{v_{k1},v_{k2},\ldots,v_{kk}\}$ and \linebreak $F_{v_{kn}}(y_{1},y_{2},\ldots,y_{n})$ represents the joint CDF of $\{v_{k1},v_{k2},\ldots,v_{kn}\}$ evaluated at $\{y_{1},y_{2},\ldots,y_{n}\}$. In fact, one can use (\ref{eq:mainresult}) for any joint greedy user scheduling and orthogonal BF algorithm with homogeneous users as long as a composite joint PDF expression can be written as in (\ref{eq:compositejointpdf}) and scheduled user's SINR at a given step is not affected from the subsequent steps.
\section{Adaptive Orthogonal Beamforming with User Selection}
\label{adaptiveOBF}
We first apply the order statistics strategy described in the previous section to adaptive OBF algorithm~\cite{Duplicy:07}. An outline of adaptive OBF is provided in Table I. From the set of $K$ users, the first user is scheduled in such a way that it has the largest channel gain (multiuser diversity) and $\overline{\textbf{h}}_{k_{1}}=\textbf{h}_{k_{1}}/\|\textbf{h}_{k_{1}}\|$ is assigned as its BF weight vector. Then, $n$th scheduled user is determined by maximizing SINR metric, which is calculated using the orthogonal projection of the candidate user's channel vector onto the null space of the previously scheduled users' BF weight vectors. The sum rate corresponding to $n$th step is denoted by $C(U_{n})$ and can be expressed as follows
\begin{displaymath}
C(U_{n})=\sum_{i=1}^{n}\log(1+\SINR_{k_{i}})
\end{displaymath}
where
\begin{displaymath}
\SINR_{k_i}=\frac{\|\textbf{h}_{k_i}\|^{2}p_{k_i}^{2}}{\|\textbf{h}_{k_i}\|^{2}(1-p_{k_i}^{2})+\frac{n}{P}}
\end{displaymath}
with  $p_{k_1}^{2}=1$ and
\begin{displaymath}
p_{k_i}^{2}=
\left\|\left(\textbf{I}_{M}-\textbf{W}(:,1:n-1)\textbf{W}^{H}(:,1:n-1)\right)\overline{\textbf{h}}_{k_i}\right\|^{2}
\end{displaymath}
for $i\geq 2$. Additionally, $\textbf{W}$ denotes the beamforming weight matrix with $\textbf{w}_{k_{i}}$ as its $i$th column and $\overline{\textbf{h}}_{k_{i}}=\textbf{h}_{k_{i}}/\|\textbf{h}_{k_{i}}\|$ denotes the normalized channel vector of $i$th scheduled user.
If we have $C(U_{n}) \leq C(U_{n-1})$ at any step, the algorithm terminates with the set of scheduled users $U_{n-1}$ and the corresponding BF weight matrix. Despite not being a part of the original algorithm in~\cite{Duplicy:07}, it allows one to adaptively choose the number of scheduled users to maximize the sum rate.
\begin{table}[!t]
\renewcommand{\arraystretch}{1.9}
\centering
\caption{Adaptive Orthogonal Beamforming with User Selection}
\vspace{-4.5mm}
\label{table1}
\begin{tabular}[t]{p{8.5cm}}
\hline
\textbf{Step 1)} Select the first user as $k_{1}\!=\!\!\underset{k \in \{1,\ldots,K\}}{\operatorname{arg\,max}}\|\textbf{h}_{k}\|^{2}$ and set $U_{1}=\{k_{1}\}$,\\
~~~~~~~~~~$\textbf{W}(:,1)=\overline{\textbf{h}}_{k_{1}}=\textbf{h}_{k_{1}}/\|\textbf{h}_{k_{1}}\|$, $n=1$, and calculate $C(U_{1})$.\\
\textbf{Step 2)} While $n<M$ \\
~~~~~~~~~~a) Increase $n$ by one and set $\SINR_{\max}=0$.\\
~~~~~~~~~~b) For $u=\{1....,K\}\setminus U_{n-1}$\\
~~~~~~~~~~~~~~~~~$p_{u}^{2}=\left\|\left(\textbf{I}_{M}-\textbf{W}(:,1:n-1)\textbf{W}^{H}(:,1:n-1)\right)\overline{\textbf{h}}_{u}\right\|^{2}$,\\
~~~~~~~~~~~~~~~~~$\SINR_{u}=\frac{\|\textbf{h}_{u}\|^{2}p_{u}^{2}}{\|\textbf{h}_{u}\|^{2}(1-p_{u}^{2})+\frac{n}{P}}$,\\
~~~~~~~~~~~~~~~~~If $\SINR_{u} > \SINR_{\max}$ \\
~~~~~~~~~~~~~~~~~~~~$\SINR_{\max}=\SINR_{u}$ and $k_{n}=u$. \\
~~~~~~~~~~c) $\textbf{W}(:,n)=\frac{\left(\textbf{I}_{M}-\textbf{W}(:,1:n-1)\textbf{W}^{H}(:,1:n-1)\right)\overline{\textbf{h}}_{k_{n}}}{\left\|
\left(\textbf{I}_{M}-\textbf{W}(:,1:n-1)\textbf{W}^{H}(:,1:n-1)\right)\overline{\textbf{h}}_{k_{n}}\right\|}$,\\ ~~~~~~~~~~~~~~$U_{n}=U_{n-1}\cup\{k_{n}\}$, and calculate $C(U_{n})$.\\
~~~~~~~~~~d) If $C(U_{n}) \leq C(U_{n-1})$, decrease $n$ by $1$ and go to step $3$. \\
\textbf{Step 3)} The set of scheduled users is $U_{n}$ and the corresponding BF \\\vspace{-0.23in}
~~~~~~~~~~weight vectors are the ordered columns of $\textbf{W}$.
\end{tabular}
\hrule
\vspace{-5mm}
\end{table}
\subsection{Performance Analysis}
In this section, we obtain PDF expressions for the scheduled users' SINR values. The number of scheduled users is a random quantity. However, for the sake of simplicity, we assume that the number of scheduled users equals $r$ with $r \in \{1,\ldots, M\}$ and denote SINR of $n$th scheduled user by $y_{n}$ where $n \in \{1,\ldots,r\}$. Note that the analysis presented here provides PDF expressions for the original adaptive OBF algorithm given in~\cite{Duplicy:07} when $r=M$. A direct approach for deriving PDFs of the scheduled users' SINRs can be calculating the PDF of the first scheduled user's SINR in the first place, and then, using conditional PDF expressions, to try to determine the PDF of the second scheduled user. However, this approach turns out to be quite complicated to generalize beyond the second step. A scheduling algorithm with interference cancellation is analyzed using this approach in~\cite{Dimic:05}. Even in this simpler case, it requires complicated algebraic operations to generalize. On the other hand, we approach the problem from a different perspective by using unordered SINRs and obtain the joint PDF of all scheduled users' SINRs at once. Define unordered SINR $v_n$ at $n$th scheduling step as the one when no user ordering is applied:
\begin{displaymath}
v_{n}=\left\{ \begin{array}{lc}
\|\textbf{h}_{1}\|^{2}\frac{P}{r}& \mbox{~~for~~}n=1,\\
\frac{\|\textbf{h}_{n}\|^{2}-\|\textbf{P}_{n}^{H}\textbf{h}_{n}\|^{2}}
{\|\textbf{P}_{n}^{H}\textbf{h}_{n}\|^{2}+\frac{r}{P}} & \mbox{~~for~~}n \in \{2,\ldots,r\},
\end{array} \right.
\end{displaymath}
where the columns of $M$-by-$(n-1)$ matrix $\textbf{P}_{n}$ $(n \geq 2)$ are BF weight vectors of previously scheduled users $(\textbf{P}_{n}^{H}\textbf{P}_{n}=\textbf{I}_{n-1})$ without ordering and do not depend on $\textbf{h}_{n}$. Since $\textbf{h}_{n}$ has an isotropic distribution, we can apply a unitary transformation on $\textbf{h}_{n}$ without changing its statistics. Assume a $M$-by-$M$ unitary matrix as $[\textbf{P}_{n}~~\mbox{null}(\textbf{P}_{n})]$ with the $M$-by-$(M-n+1)$ matrix $\mbox{null}(\textbf{P})$ representing the null space of $\textbf{P}$ \footnote{The matrix $\mbox{null}(\textbf{P}_{n})$ has orthonormal columns that complement the subspace spanned by the columns of $\textbf{P}$.}. If we apply this transformation above~\cite{Sharma:05}, we get
\begin{equation}
\left\|\textbf{P}_{n}^{H}~\left[\textbf{P}_{n} ~~\mbox{null}(\textbf{P}_{n})\right]~ \textbf{h}_{n}\right\|^{2}=\left\|\left[\textbf{I}_{n-1} ~~\textbf{0}_{n-1\times M-n+1}\right]~ \textbf{h}_{n}\right\|^{2}.
\label{eq:transformation1}
\end{equation}
Let $\{x_{1},\ldots,x_{r}\}$ be independent random variables with $\{x_{1},\ldots,x_{r-1}\}$ being exponentially distributed random variables and $x_{r}$ denoting a chi-squared random variable with $2(M-r+1)$ degrees of freedom. Then, using (\ref{eq:transformation1}), we can write unordered SINRs as
\begin{equation}
v_{n}=\left\{ \begin{array}{lc}
\left(x_{1}+\ldots+x_{r}\right)\frac{P}{r} & \mbox{ if  } n=1,\\
\frac{x_{n}+\ldots+x_{r}}{x_{1}+\ldots+x_{n-1}+\frac{r}{P}} & \mbox{otherwise}.
       \end{array} \right.
       \label{eq:equnorderedSINRs}
\end{equation}
In (\ref{eq:equnorderedSINRs}), $\{x_{1},\ldots,x_{r}\}$ have the following joint PDF
\begin{equation}
f(x_{1},\ldots,x_{r})=e^{-\left(x_{1}+\ldots+x_{r}\right)}\frac{x_{r}^{M-r}}{\Gamma(M-r+1)}
\label{eq:unorderedjointpdf}
\end{equation}
for $\{x_{1},\ldots,x_{r}\} \geq 0$ where $\Gamma(s)$ denotes the gamma function~\cite{Gradshteyn:00}. We need the joint PDF of unordered SINRs. Thus, with the help of (\ref{eq:equnorderedSINRs}), we can derive the following transformation:
\begin{equation}
x_{n}\!=\!\left\{ \begin{array}{lc}
\!\!\! \lambda_n\!=\frac{r}{P}(1+v_{1})\frac{v_{n}-v_{n+1}}{(1+v_{n})(1+v_{n+1})} & \!\!\!\! \mbox{for~} n\! \in \! \{1,\ldots,r-1\},\\
\!\!\! \lambda_r=\frac{r}{P}(1+v_{1})\frac{v_{r}}{1+v_{r}} & \mbox{for~} n=r,
       \end{array} \right.
       \label{eq:eqxssolved}
\end{equation}
with the Jacobian determinant given by
\begin{equation}\det(J)=
\left(\frac{r}{P}\right)^{r}\frac{\left(1+v_1\right)^{r-1}}{\prod_{k=2}^{r}\left(1+v_k\right)^{2}}.
\end{equation}
Consequently, the joint PDF of $\{v_{1},\ldots,v_{r}\}$ can be written as~\cite{Papoulis:02}
\begin{flalign}
f(v_{1},\ldots,v_{r})=f(x_{1}=\lambda_1,\ldots,x_{r}=\lambda_r)~|\det(J)| && \nonumber
\end{flalign}
\begin{flalign}
&& =\left(\frac{r}{P}\right)^{M}\frac{e^{-v_{1}\frac{r}{P}}}{\Gamma(M-r+1)}
\frac{\left(1+v_1\right)^{M-1}}{\prod_{k=2}
^{r}\left(1+v_k\right)^{2}}
\left(\frac{v_r}{1+v_r}\right)^{M-r}
\end{flalign}
for $v_1\geq v_2\geq \ldots\geq v_{r} \geq 0$. The joint PDF of $\{v_1,\ldots,v_n\}$ for $n \in \{1,\ldots,r\}$ can be found by integrating out $\{v_{n+1},\ldots,v_r\}$ as
\begin{flalign}
f(v_{1},\ldots,v_{n}) && \nonumber
\end{flalign}
\vspace{-.25in}
\begin{flalign}
=\int_{0}^{v_n}\ldots\int_{0}^{v_{r-2}}\int_{0}^{v_{r-1}}f(v_{1},\ldots,v_{r}) dv_{r}dv_{r-1}\ldots dv_{n+1} && \nonumber
\end{flalign}
\vspace{-.15in}
\begin{flalign}
=\left(\frac{r}{P}\right)^{M}\frac{e^{-v_{1}\frac{r}{P}}}{\Gamma(M-n+1)}\frac{\left(1+v_1\right)^{M-1}}{\prod_{k=2}
^{n}\left(1+v_k\right)^{2}}
\left(\frac{v_n}{1+v_n}\right)^{M-n}. &&
\label{eq:jointpdfofv}
\end{flalign}
Note that the preceding expression is the joint PDF of the resulting SINRs when the user selection algorithm in Table I is replaced by a random user selection technique. In order to obtain the PDF expressions of the scheduled users, we make use of the following theorem~\cite{Ozyurt:11}. This enables us to succinctly write the SINR of $n$th scheduled user, i.e., $y_n$, as a function of unordered SINRs.
\begin{figure*}[!hb]
\hrule
\addtocounter{equation}{4}
\vspace{.075in}
\begin{eqnarray}
\varphi_{3}(y_3,y_2,y_1)&=&\int_{y_3}^{y_2}\int_{v_2}^{y_1}f(v_1,v_2,v_3=y_3)dv_{1}dv_{2} \nonumber \\
&=&\frac{\left(\frac{r}{P}\right)^{M}}{\Gamma(M-2)}\frac{y_{3}^{M-3}}{(1+y_{3})^{M-1}}\int_{y_3}^{y_2}\int_{v_2}^{y_1}e^{-v_{1}\frac{r}{P}}
\frac{(1+v_1)^{M-1}}{(1+v_2)^2}dv_{1}dv_{2}  \nonumber \\
&=&\frac{e^{\frac{r}{P}}}{\Gamma(M-2)}\frac{y_{3}^{M-3}}{(1+y_{3})^{M-1}}\left\{\frac{\Gamma\left(M,\frac{r}{P}(1+y_3)\right)}{1+y_3}\right.-
\frac{\Gamma\left(M,\frac{r}{P}(1+y_2)\right)}{1+y_2} \label{eq:varphi321} \\
&~&~~-\frac{y_2-y_3}{(1+y_2)(1+y_3)}\Gamma\left(M,\frac{r}{P}(1+y_1)\right)+\frac{r}{P}\Bigg[\Gamma\left(M-1,\frac{r}{P}(1+y_2)\right)-\Gamma\left(M-1,\frac{r}{P}(1+y_3)\right)\Bigg]\Bigg\}.
\nonumber
\end{eqnarray}
\vspace{.075in}
\hrule
\vspace{.075in}
\begin{eqnarray}
I_{3}(y_3,y_2,y_1)&=&\int_{0}^{y_3}\varphi_{3}(\alpha,y_2,y_1)d\alpha \nonumber \\
&=&\frac{e^{\frac{r}{P}}}{\Gamma(M-2)}\sum_{i=0}^{M-3}{M-3 \choose i}(-1)^{i}\Bigg\{\frac{(1+y_{3})^{i+1}-1}{(1+y_{3})^{i+1}(i+1)}\Bigg[\frac{r}{P}\Gamma\left(M-1,\frac{r}{P}(1+y_{2})\right) \nonumber \\
&~&~~+\frac{\Gamma
\left(M,\frac{r}{P}(1+y_{1})\right)-\Gamma\left(M,\frac{r}{P}(1+y_{2})\right)}
{1+y_{2}}\Bigg]-\frac
{(1+y_{3})^{i+2}-1}{(1+y_{3})^{i+2}(i+2)}~\Gamma\left(M,\frac{r}{P}(1+y_{1})
\right) \nonumber \\
&~&~~+\Bigg[\frac{r}{P}\frac{\Gamma\left(M-1,\frac{r }{P}(1+y_{3})\right)}
{(1+y_{3})^{i+1}(i+1)}-\frac{\Gamma\left(M,\frac{r}{P}(1+y_{3})\right)}
{(1+y_{3})^{i+2}(i+2)}-\left(\frac{r}{P}\right)^{i+2} \frac{\Gamma
\left(M-i-2,\frac{r}{P}(1+y_{3})\right)}{(i+1)(i+2)}\Bigg] \nonumber \\
&~&~~-\Bigg[ \frac{r}{P}\frac{\Gamma\left(M-1,\frac{r}{P}\right)}{(i+1)}-\frac{ \Gamma\left(M,\frac{r}{P}\right)}{i+2}-\left(\frac{r}{P}\right)^{i+2}
\frac{\Gamma\left(M-i-2,\frac{r}{P}\right)}{(i+1) (i+2)}\Bigg]\Bigg\}.
\label{eq:I321}
\end{eqnarray}
\addtocounter{equation}{-6}
\end{figure*}
\newtheorem{theorem}{Theorem}
\begin{theorem}\label{thm1:th}
The joint PDF of $\{y_1,y_2,\ldots,y_n\}$ can be expressed as
\begin{flalign}
f(y_1,\ldots,y_n)=\frac{K!}{(K-n)!} && \nonumber
\end{flalign}
\vspace{-.1in}
\begin{equation}
\times \left(\int_{0}^{y_n}\varphi_{n}(\alpha,y_{n-1},\ldots,y_1)d\alpha\right)^{K-n} ~~ \prod_{i=1}^{n}\varphi_{i}(y_i,\ldots,y_1)
\label{eq:teorem1}
\end{equation}
for $y_1 \geq \ldots \geq y_n \geq 0$ and $n \in \{1,\ldots,r\}$. In (\ref{eq:teorem1}), we have
\begin{equation}
\varphi_{1}(y_1)=f(v_1=y_1)
\label{eq:varphi1}
\end{equation}
and
\begin{flalign}
\varphi_{n}(y_n,y_{n-1},\ldots,y_1) && \nonumber
\end{flalign}
\vspace{-.2in}
\begin{flalign}
=\int_{y_n}^{y_{n-1}}\int_{v_{n-1}}^{y_{n-2}}\ldots\int_{v_2}^{y_1}f(v_1,v_2,\ldots,v_{n-1},v_n=y_n) && \nonumber
\end{flalign}
\vspace{-.25in}
\begin{equation}
\hspace{1.65in} \times dv_{1}\ldots dv_{n-2} dv_{n-1}
\label{eq:varphin}
\end{equation}
which can be evaluated in closed-form for any $n$. A similar proposition is made in~\cite{Dimic:05} with a lengthy proof. It can be considered as a special case of the result given in Section~\ref{mainresult}.
\end{theorem}
\begin{IEEEproof}
See Appendix~\ref{appendixA}.
\end{IEEEproof}
The marginal PDF of $y_n$ can be obtained as
\begin{displaymath}
f(y_n)=\int_{y_{n}}^{\infty}\ldots \int_{y_{3}}^{\infty}\int_{y_{2}}^{\infty}f(y_1,y_2,\ldots,y_n)dy_{1}dy_{2}\ldots dy_{n-1}
\end{displaymath}
for $y_n \geq 0$. The average sum rate can be calculated as
\begin{displaymath}
\sum_{n=1}^{r}\mathds{E}\left[\log\left(1+y_n\right)\right]=\sum_{n=1}^{r}\int_{0}^{\infty}\log\left(1+y_n\right)f(y_n)dy_n.
\end{displaymath}
\subsubsection{Example}
While the above theorem and the joint PDF of unordered SINRs in (\ref{eq:jointpdfofv}) can be applied for any number of scheduled users, we demonstrate the application of the tools developed above for the first three scheduled users. Let us assume that $r$ out of $K$ users are scheduled with\linebreak $\{K, M \} \geq r \geq 3$. Evaluating (\ref{eq:varphi1}) and (\ref{eq:varphin}) for $n=1$ and $n=2$, we get
\begin{displaymath}
\varphi_{1}(y_1)=f(v_1 = y_1)=\left(\frac{r}{P}\right)^{M}
\frac{y_{1}^{M-1}}{\Gamma(M)}e^{-y_{1}\frac{r}{P}}
\end{displaymath}
and
\begin{flalign}
\varphi_{2}(y_2,y_1)=\int_{y_2}^{y_1}f(v_1,v_2=y_2)dv_{1} && \nonumber
\end{flalign}
\vspace{-.15in}
\begin{flalign}
&& =e^{\frac{r}{P}}~y_{2}^{M-2}~\frac{\Gamma\left(M,\frac{r}{P}(1+y_2)\right)-\Gamma\left(M,\frac{r}{P}(1+y_1)\right)}{\Gamma(M-1)(1+y_2)^{M}}
\nonumber
\end{flalign}
where $\Gamma(s,x)$ is the upper incomplete gamma function given by
\begin{equation}
\Gamma(s,x)=\Gamma(s)e^{-x}\sum_{i=0}^{s-1}\frac{x^{i}}{i!}
\label{eq:gammafonksiyonu}
\end{equation}
for positive integer $s$ values~\cite{Gradshteyn:00}.
\begin{figure}[!t]
\centering
\vspace{-.075in}
\includegraphics[width=0.53\textwidth]{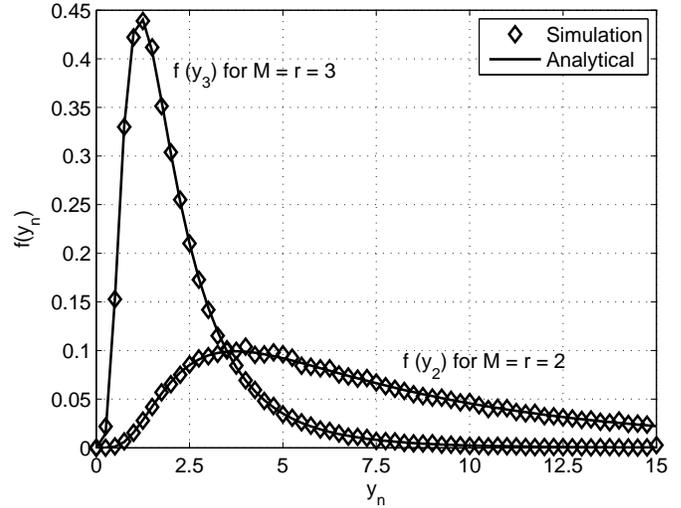}
\vspace{-3mm}
\caption{Comparison of the analytical PDF expressions with the numerical results for adaptive OBF with $P=15$ dB and $K=10$.}
\label{OrtBF_Fig1:fig}
\end{figure}
We can write $\varphi_3(y_3,y_2,y_1)$ in closed-form as given in (\ref{eq:varphi321}) as a double-column equation. Defining $I_3(y_3,y_2,y_1)=\int_{0}^{y_3}\varphi_{3}(\alpha,y_2,y_1)d\alpha$, one can obtain (\ref{eq:I321}) at the bottom of this page where the derivation is based on the binomial expansion formula~\cite{Gradshteyn:00} and omitted for the simplicity. Finally, the joint PDF of the first three scheduled users' SINRs can be written as
\addtocounter{equation}{2}
\begin{flalign}
f(y_{1},y_{2},y_{3})=\frac{K!}{(K-3)!}I_{3}(y_3,y_2,y_1)
^{K-3} && \nonumber
\end{flalign}
\begin{equation}
\hspace{.65in} \times \varphi_{1}(y_1)\varphi_{2}(y_2,y_1)\varphi_{3}(y_3,y_2,y_1)
\end{equation}
for $y_1 \geq y_2 \geq y_3 \geq 0$. The closed-form expressions for the marginal PDFs of $y_1$ and $y_2$ can be found in~\cite{Ozyurt:11} and the marginal PDF of $y_3$ can be obtained either analytically or using numerical integration techniques.

The analytically obtained PDF expressions are plotted in Fig.~\ref{OrtBF_Fig1:fig} together with the corresponding simulated histograms for $P=15$ dB and $K=10$. Two different $M$ values, namely $M=\{2,3\}$, are used with $r=M$ and the simulated results are based on $10^5$ realizations. The strong match between analytical and simulated results verifies the accuracy of the analytical derivation.
\section{Orthogonal Linear Beamforming with User Selection}
\label{OLBF}
Another orthogonal BF algorithm (OLBF) has been proposed in~\cite{Francisco:07} and shown to achieve similar performance in low SNR regime with a lower complexity as compared to adaptive OBF. Akin to adaptive OBF, OLBF schedules the first user in such a way that it has the largest channel gain and $\overline{\textbf{h}}_{k_{1}}=\textbf{h}_{k_{1}}/\|\textbf{h}_{k_{1}}\|$ is assigned as its BF weight vector. Then, applying Gram-Schmidt orthogonalization process to $\overline{\textbf{h}}_{k_{1}}$, $(M-1)$ orthonormal BF weight vectors are obtained. Assignment of these $(M-1)$ BF weight vectors to users is outlined in Table II. For each BF weight vector, the user with highest SINR is scheduled from the set of candidate users in a sequential manner. Note that when the number of scheduled users is less than $M$ where the set of BF weight vectors does not span $M$ dimensional space, the SINR expression used in Table II becomes a lower bound for the original SINR. Hence, in the following analysis, we assume that the number of scheduled users is equal to $M$ $(r=M)$ as given for the original OLBF algorithm in~\cite{Francisco:07}.
\subsection{Performance Analysis}
In this section, the result of Section~\ref{mainresult} is used for statistical analysis of OLBF algorithm. Integral evaluations are mainly based on the binomial expansion formula~\cite{Gradshteyn:00} and the expansion of the upper incomplete gamma function in (\ref{eq:gammafonksiyonu}), hence omitted for the sake of simplicity. Similar to Section~\ref{adaptiveOBF}, we define unordered SINR $v_n$ at $n$th scheduling step as the one when no user ordering is applied. From Table II, one can write
\begin{displaymath}
v_{n}=\left\{ \begin{array}{lc}
\|\textbf{h}_{1}\|^{2}\frac{P}{M} & \mbox{~for~} n=1,\\
\frac{\|\textbf{h}_{n}\|^{2}-\|\textbf{P}_{n}^{H}\textbf{h}_{n}\|^{2}}
{\|\textbf{P}_{n}^{H}\textbf{h}_{n}\|^{2}+\frac{M}{P}} & \mbox{~for~} n \in \{2,\ldots, M\},
       \end{array} \right.
       \end{displaymath}
where the columns of $M$-by-$(M-1)$ matrix $\textbf{P}_{n}$ are BF weight vectors of $(M-1)$ scheduled users other than $n$th one without user ordering $(\textbf{P}_{M-1}^{H}\textbf{P}_{M-1}=\textbf{I}_{M-1})$. Applying a unitary transformation as $[\textbf{P}_{n}~~\mbox{null}(\textbf{P}_{n})]$ on $\textbf{h}_{n}$ above (without affecting its statistics), we can write
\begin{equation}
\left\|\textbf{P}_{n}^{H}~\left[\textbf{P}_{n} ~~\mbox{null}(\textbf{P}_{n})\right]~ \textbf{h}_{n}\right\|^{2}=\left\|\left[\textbf{I}_{M-1} ~~\textbf{0}_{M-1\times 1}\right]~ \textbf{h}_{n}\right\|^{2}.
\label{eq:transformation2}
\end{equation}
Defining $\{x_{1},\ldots,x_{M}\}$ as IID exponentially distributed random variables and using (\ref{eq:transformation2}), we can write
\begin{equation}
v_{n}=\left\{ \begin{array}{lc}
\left(x_{1}+\ldots+x_{M}\right)\frac{P}{M} & \mbox{ for  } n=1,\\
\frac{x_{n}}{x_{1}+\ldots+x_{n-1}+x_{n+1}+\ldots+x_{M}+\frac{M}{P}} & \mbox{\!for } n \in \{2,\ldots, M\}.
       \end{array} \right.
       \end{equation}
The joint PDF of $\{x_{1},\ldots,x_{M}\}$ is given by
\begin{equation}
f(x_{1},\ldots,x_{M})=e^{-\left(x_{1}+\ldots+x_{M}\right)}
\end{equation}
for $\{x_{1},\ldots,x_{M}\} \geq 0$. Solving for $x_{n}$, we get
\begin{equation}
x_{n}=\left\{ \begin{array}{lc}
\!\! \lambda_1=\frac{M}{P}v_{1}-\frac{M}{P}(1+v_1)\sum_{k=2}^{M}\frac{v_k}{1+v_k}& \mbox{~for~} n=1,\\
\!\! \lambda_n=\frac{M}{P}(1+v_1)\frac{v_n}{1+v_n} & \hspace{-.4in} \mbox{for~} n \in \{2,\ldots, M\},
       \end{array} \right.
       \end{equation}
with the following Jacobian determinant
\begin{equation}\det(J)=
\left(\frac{M}{P}\right)^{M}\frac{\left(1+v_1\right)^{M-1}}{\prod_{k=2}^{M}
\left(1+v_k\right)^{2}}.
\end{equation}
Consequently, the joint PDF of $\{v_{1},\ldots,v_{M}\}$ can be written as~\cite{Papoulis:02}
\begin{eqnarray}
f(v_{1},\ldots,v_{M})&=&f(x_{1}=\lambda_{1},\ldots,x_{M}=\lambda_{M})~|\det(J)|
\nonumber \\
&=&\left(\frac{M}{P}\right)^{M}e^{-v_{1}\frac{M}{P}}
\frac{\left(1+v_1\right)^{M-1}}{\prod_{k=2}
^{M}\left(1+v_k\right)^{2}}
\label{eq:unorderedjointpdfOLBF}
\end{eqnarray}
for $v_1\geq \{v_2,\ldots,v_{M}\} \geq 0$ and $\frac{v_1}{1+v_1}\geq \sum_{k=2}^{M}\frac{v_k}{1+v_k}$. For more than two scheduled users, the approach used for the statistical analysis of adaptive OBF in Section~\ref{adaptiveOBF} cannot be directly applied in the same manner for OLBF. Difficulty comes from the fact that the sorting among the random variables in (\ref{eq:unorderedjointpdfOLBF}) is not simple and it is quite involved to evaluate the CDF $F_{v_{M}}(v_{1},v_{2},\ldots,v_{M})$ at $\{y_1, y_2,\ldots,y_M\}$ given in (\ref{eq:mainresult}). For instance, the sorting rule can be written as
\begin{displaymath}\left\{
{\begin{array}{c}
v_{1} \geq v_3 \geq 0,\\
\frac{v_{1}-v_{3}}{1+2 v_{3}+v_{1}v_{3}} \geq  v_2 \geq 0,
\end{array}} \right.
\end{displaymath}
when $M=3$.
\begin{table}[!t]
\renewcommand{\arraystretch}{1.9}
\centering
\caption{Orthogonal Linear Beamforming with User Selection}
\vspace{-4.5mm}
\label{table2}
\begin{tabular}[t]{p{8.5cm}}
\hline
\textbf{Step 1)} Select the first user as $k_{1}\!=\!\!\underset{k \in \{1,\ldots,K\}}{\operatorname{arg\,max}}\|\textbf{h}_{k}\|^{2}$ and set $U_{1}=\{k_{1}\}$,\\
~~~~~~~~~~$\overline{\textbf{h}}_{k_{1}}=\textbf{h}_{k_{1}}/\|\textbf{h}_{k_{1}}\|$, and $n=1$.\\
\textbf{Step 2)} Compute $(M-1)$ orthonormal basis vectors spanning the null\\
~~~~~~~~~~  space of $\overline{\textbf{h}}_{k_{1}}$ and form a matrix $\textbf{W}_{2}$ as these basis vectors being\\
~~~~~~~~~~  its columns.\\
\textbf{Step 3)} While $n<M$ \\
~~~~~~~~~~a) Increase $n$ by one and set $\SINR_{\max}=0$.\\
~~~~~~~~~~b) For $u=\{1....,K\}\setminus U_{n-1}$\\
~~~~~~~~~~~~~~~~~$q_{u}^{2}=\left|\textbf{W}_{2}^{H}(:,n-1)\overline{\textbf{h}}_{u}\right|^{2}$,\\
~~~~~~~~~~~~~~~~~$\SINR_{u}=\frac{\|\textbf{h}_{u}\|^{2}q_{u}^{2}}{\|\textbf{h}_{u}\|^{2}(1-q_{u}^{2})+\frac{M}{P}}$,\\
~~~~~~~~~~~~~~~~~If $\SINR_{u} > \SINR_{\max}$ \\
~~~~~~~~~~~~~~~~~~~~$\SINR_{\max}=\SINR_{u}$ and $k_{n}=u$.\\
~~~~~~~~~~c) $U_{n}=U_{n-1}\cup\{k_{n}\}$.\\
\textbf{Step 4)} The set of scheduled users is $U_{M}$ and the corresponding BF \\\vspace{-0.23in}
~~~~~~~~~~weight vectors are the ordered columns of $\textbf{W}=[\overline{\textbf{h}}_{k_{1}}~\textbf{W}_2]$.
\end{tabular}
\hrule
\vspace{-3.5mm}
\end{table}
Even in this case, it is cumbersome to analytically evaluate $F_{v_3}(v_1,v_2,v_3)$ at $\{y_1,y_2,y_3\}$. In order to address this issue, we apply a one-to-one transformation on each unordered SINR to obtain a simpler sorting rule among the variables. Representing these new variables by $z_n$ for $n \in \{1,\ldots,M\}$, the transformation from $v_n$ to $z_n$ is given by
\begin{equation}
z_n=\frac{v_n}{v_{n}+1}.
\label{eq:vn_to_zn}
\end{equation}
With this transformation, the new sorting rule can be written as
\begin{displaymath}\left\{
{\begin{array}{c}
0 \leq z_n \leq 1,\\
0 \leq z_2+z_3+\ldots+z_M \leq z_1 \leq 1,
\end{array}} \right.
\end{displaymath}
for all $n$. Using (\ref{eq:unorderedjointpdfOLBF}) and (\ref{eq:vn_to_zn}), the joint PDF of $\{z_1,z_2,\ldots,z_M\}$ can be found to be~\cite{Papoulis:02}
\begin{equation}
f(z_{1},z_{2},\ldots,z_M)=e^{-\frac{M}{P}\frac{z_{1}}{1-z_{1}}}\left(\frac{M}{P}\right)^{M}\frac{1}{(1-z_{1})^{M+1}}.
\label{eq:jointpdfofzM}
\end{equation}
\vspace{.5in}
Similarly, the joint PDF of $\{z_1,z_2,\ldots,z_n\}$ can be found to be
\begin{equation}
f(z_{1},z_{2},\ldots,z_n)=\frac{e^{-\frac{M}{P}\frac{z_{1}}{1-z_{1}}}(M/P)^{M}}{(1-z_{1})^{M+1}}\frac{\left(z_{1}
-\sum_{i=2}^{n}z_{i}\right)^{M-n}}{(M-n)!}
\label{eq:jointpdfofzn}
\end{equation}
for $0 \leq z_2+\ldots+z_{n} \leq z_1 \leq 1$, $0 \leq z_n \leq 1$, and $n \in \{1,2,\ldots,M\}$. We represent transformed SINR of $n$th scheduled user by $t_{n}$ when the corresponding unordered SINR expression is given by $z_n$. Hence, the original SINR of $n$th scheduled user is given by $y_n=t_{n}/(1-t_{n})$ and can be obtained by a simple change of variables. The following theorem provides a compact solution on the joint PDF of $\{t_{1}, t_{2}, \ldots, t_{n}\}$.
\begin{theorem}\label{thm2:th}
The joint PDF of $\{t_{1}, t_{2}, \ldots, t_{n}\}$ with\linebreak $n \in \{1,2,\ldots,M\}$ can be written as
\begin{flalign}
f(t_{1},t_{2},\ldots,t_{n})=\frac{K!}{(K-n)!} && \nonumber
\end{flalign}
\begin{equation}
\times
\left(F_{z_{n}}(t_1,t_2,\ldots,t_n)
\right)^{K-n}\prod_{k=1}^{n}
\xi_{k}(t_k,t_{k-1},\ldots,t_1)
\label{eq:teorem2}
\end{equation}
with $0 \leq \{t_{2},\ldots,t_{n}\} \leq t_1 \leq 1$. In (\ref{eq:teorem2}), $F_{z_{n}}(t_1,t_2,\ldots,t_n)$ is the CDF of $\{z_1,z_2,\ldots,z_n\}$ evaluated at $\{t_1,t_2,\ldots,t_n\}$, $\xi_1(t_1)=f(z_1=t_1)$, and
\begin{flalign}
\xi_{k}(t_k,t_{k-1},\ldots,t_1) && \nonumber
\end{flalign}
\begin{displaymath}
=\int_{R_{\xi_k}} f(z_1,z_2,\ldots,z_{k-1},z_k=t_k) dz_{1}dz_{2}\ldots dz_{k-1}
\end{displaymath}
where $f(z_1,z_2,\ldots,z_k)$ is the joint PDF of $\{z_1,z_2,\ldots,z_k\}$ with $k \in \{1,2, \ldots, n\}$. The integration region $R_{\xi_k}$ is given by union of the regions represented by
\begin{equation} \left\{
{\begin{array}{c}
0 \leq z_2+z_3+\ldots+ z_{k-1}+t_k \leq z_1 \leq 1,\\
0 \leq z_n \leq t_n \leq 1 \text{~~for~~} n \in \{1,2, \ldots, k-1\}.
\end{array}} \right.
\label{eq:Dir_region}
\end{equation}
\end{theorem}
\begin{IEEEproof}
This theorem can be directly obtained from the result of Section~\ref{mainresult}. For transformed SINRs defined above, the PDF expressions in (\ref{eq:compositejointpdf}) and the CDF expressions in (\ref{eq:mainresult}) are independent of $k$. Combining the CDFs into a single term and applying a change as $\phi_{k}(t_k,t_{k-1},\ldots,t_1)\rightarrow\xi_{k}(t_k,t_{k-1},\ldots,t_1)$ in (\ref{eq:mainresult}), the result given in (\ref{eq:teorem2}) can be obtained.
\end{IEEEproof}

\begin{figure}[!t]
        \centering
        \resizebox{.475\textwidth}{!}{\input{save2.pstex_t}}
        \vspace{.1in}
                \caption{Illustration of the inclusion-exclusion principle for $n=3$ where the shaded areas refer to integration regions.}
   \label{OrtBF_InExprinciple:fig}
    \end{figure}
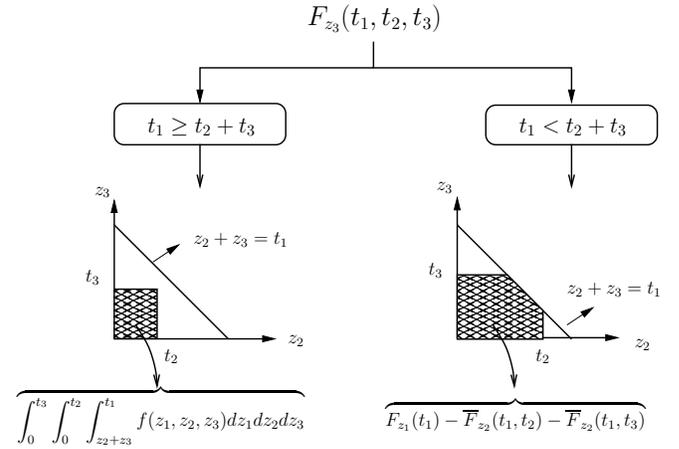

In order to utilize Theorem~\ref{thm2:th}, we need to express CDF of $\{z_1,z_2,\ldots,z_n\}$. This has multiple segmentations for $n \geq 3$ and can pose high complexity for higher values of $n$. In fact, when $n=4$, there are eleven segments that one needs to consider separately. Here, we note the similarity of the region the random variables in (\ref{eq:jointpdfofzn}) are defined to a Dirichlet distribution~\cite{Ng:11}. In~\cite{Gouda:10}, the high complexity involved in evaluation of a high dimensional multivariate CDF is addressed in a Dirichlet distribution context. Using the inclusion-exclusion principle~\cite{David:03} with an earlier result on multivariate Dirichlet distribution~\cite{Szantai:85}, the authors develop a recursive technique that allows the high dimensional multivariate CDF expression to be represented by marginal CDFs of individual variables. In the following, we apply the technique described in~\cite{Gouda:10} with some manipulation for the calculation of $F_{z_{n}}(t_1,t_2,\ldots,t_n)$. When $t_1 \geq t_2 + t_3 + \ldots + t_n$, $F_{z_{n}}(t_1,t_2,\ldots,t_n)$ can be calculated for any $n$ value and the derivation is deferred to Appendix~\ref{appendixB}. Let the survival function~\cite{Ng:11} of $\{z_1,z_2, \ldots,z_n\}$ be defined as $\overline{F}_{z_{n}}(t_1,t_2,\ldots,t_n)=\text{Pr}~(z_1 \leq t_1,z_2 > t_2,\ldots,z_n > t_n)$. It is worth to mention that this survival function is not the complementary CDF of $\{z_1,z_2, \ldots,z_n\}$ evaluated at $\{t_1,t_2,\ldots,t_n\}$. Assuming $0 \leq \{t_1,t_2,\ldots, t_n\} \leq 1$, we first calculate the CDFs of $z_1$ and $\{z_1,z_2\}$ together with the survival function of $\{z_1,z_2\}$ as
\begin{eqnarray}
F_{z_{1}}(t_1)&=&\int_{0}^{t_1} f(z_1) dz_1, \nonumber\\
F_{z_{2}}(t_1,t_2)&=&\int_{0}^{t_2}\int_{z_2}^{t_1} f(z_1,z_2) dz_1 dz_2, \nonumber
\end{eqnarray}
and $\overline{F}_{z_{2}}(t_1,t_2)=F_{z_{1}}(t_1)-F_{z_{2}}(t_1,t_2)$. Second, we obtain the CDF and survival function of $\{z_1,z_2,z_3\}$ as
\begin{flalign}
F_{z_{3}}(t_1,t_2,t_3) && \nonumber
\end{flalign}
\begin{equation}
=\left\{ \begin{array}{lc}
\mbox{Given in Appendix~\ref{appendixB} ~} & \mbox{ for  } t_1 \geq t_2 + t_3,\\
F_{z_{1}}(t_1)-\overline{F}_{z_{2}}(t_1,t_2)-\overline{F}_{z_{2}}(t_1,t_3) & \mbox{ for  } t_1 < t_2 + t_3,
       \end{array} \right.
       \label{eq:CDFz3}
\end{equation}
and
\begin{flalign}
\overline{F}_{z_{3}}(t_1,t_2,t_3) && \nonumber
\end{flalign}
\begin{equation}
=\left\{ \begin{array}{l}
F_{z_{1}}(t_1)-F_{z_{2}}(t_1,t_2)-F_{z_{2}}(t_1,t_3)+F_{z_{3}}(t_1,t_2,t_3) \vspace{.05in} \\
\hspace{2in}\mbox{ ~for }
 t_1 \geq t_2 + t_3,  \vspace{.1in} \\
0 \hspace{2in} \mbox{ for  } t_1 < t_2 + t_3,
       \end{array} \right.
\end{equation}
respectively. Using the fact that $z_2 +z_3 \leq z_1$ from (\ref{eq:jointpdfofzn}), $\overline{F}_{z_{3}}(t_1,t_2,t_3)=\text{Pr}~(z_1 \leq t_1,z_2 > t_2,z_3 > t_3)$ can be determined to be equal to zero when $t_1 < t_2 + t_3$. In Fig.~\ref{OrtBF_InExprinciple:fig}, the inclusion-exclusion principle used in (\ref{eq:CDFz3}) is illustrated for $n = 3$. The shaded areas represent the integration regions for the evaluation of the CDF at $\{t_1,t_2,t_3\}$.
Proceeding same way as above, we can compute $F_{z_{n}}(t_1,t_2,\ldots,t_n)$ for any $n$ with $n \in \{3, \ldots, M \}$ in a recursive manner as given in (\ref{eq:F_zn}) on the next page~\cite{Ng:11, Gouda:10}.
\begin{figure*}[!hb]
\hrule
\begin{eqnarray}
F_{z_{n}}(t_1,t_2,\ldots,t_n)&=&\left\{ \begin{array}{l}
\mbox{Given in Appendix~\ref{appendixB}}~~~~~~~~~~~~~~~~~~~~~~~\mbox{~~~~for~~} t_1 \geq t_2 + t_3+\ldots+t_n ,\\
F_{z_{1}}(t_1)-\sum_{i=2}^{n}\overline{F}_{z_{2}}(t_1,t_i)\\
+\sum_{i=2}^{n-1}\sum_{j=i+1}^{n}\overline{F}_{z_{3}}(t_1,t_i,t_j)\\
\vdots\\
+(-1)^{n}\sum\limits_{\substack{2 \leq i_{1} < i_{2} <\\ \ldots <i_{n-2} \leq n}}\overline{F}_{z_{n-1}}(t_1,t_{i_1},\ldots,t_{i_{n-2}})
  \mbox{~~for~~} t_1 < t_2 + t_3+\ldots+t_n ,
       \end{array} \right.
       \label{eq:F_zn} \\
&~& \nonumber \\
&~& \nonumber \\
\overline{F}_{z_{n}}(t_1,t_2,\ldots,t_n)&=&\left\{ \begin{array}{lc}
F_{z_{1}}(t_1)-\sum_{i=2}^{n}F_{z_{2}}(t_1,t_i)\\
+\sum_{i=2}^{n-1}\sum_{j=i+1}^{n}F_{z_{3}}(t_1,t_i,t_j)\\
\vdots\\
+(-1)^{n-1}F_{z_{n}}(t_1,t_{2},\ldots,t_{n})
 & \mbox{~~for~~} t_1 \geq t_2 + t_3+\ldots+t_n ,\\
0 & \mbox{~~for~~} t_1 < t_2 + t_3+\ldots+t_n.
       \end{array} \right.
\end{eqnarray}
\hrule
\begin{eqnarray}
\eta(x) &=& \int_{0}^{x}\int_{z_{2}+t_3}^{t_1} f(z_1,z_2,z_3=t_3) dz_1 dz_2 \nonumber \\
&=&\sum _{i=0}^{M-3} {M-3 \choose i}\frac{(-1)^i}{\Gamma(M-2)}\left(\frac{M}{P}\right)^i e^{M/P}
\Bigg\{
-\Gamma\left(M-i,\frac{M}{P(1-t_1)}\right)\frac{(1-t_3)^{M-i-2}-(1-x-t_3)
^{M-i-2}}{M-i-2} \nonumber \\
&~& ~~~+\Gamma(M-i)\left(\frac{M}{P}\right)^{M-i-2}~~\sum _{j=0}^{M-i-1} ~\frac{\Gamma\left(i+j+2-M,\frac{M}{P(1- t_3)}\right)-\Gamma\left(i+j+2-M,\frac{M}{P (1-x-t_3)}\right)}{j!}\Bigg\}.
\label{eq:eta_x}
\end{eqnarray}
\hrule
\addtocounter{equation}{2}
       \begin{eqnarray}
F_{z_2}(t_1,t_2)&=&\int_{0}^{t_2}\int_{z_2}^{t_1}f(z_1,z_2)dz_1 dz_2  \nonumber \\
&=&\sum_{i=0}^{M-2}{M-2 \choose i}\frac{(-1)^i}{\Gamma(M-1)}\left(\frac{M}{P}\right)^{i}e^{M/P}
\Bigg\{-
\Gamma\left(M-i,\frac{M}{P(1-t_1)}\right)\frac{1-(1-t_2)^{M-i-1}}{M-i-1} \nonumber \\
&~& ~~~~+\Gamma(M-i)\left(\frac{M}{P}\right)^{M-i-1}~~ \sum_{j=0}^{M-i-1} ~ \frac{\Gamma\left(
i+j+1-M,\frac{M}{P}\right)-\Gamma\left(
i+j+1-M,\frac{M}{P(1-t_2)}\right)}{j!}\Bigg\}.
\label{eq:F_z2}
\end{eqnarray}
\end{figure*}
\addtocounter{equation}{-3}
Theoretically, one can calculate $F_{z_{n}}(t_1,t_2,\ldots,t_n)$ and $\overline{F}_{z_{n}}(t_1,t_2,\ldots,t_n)$ for any dimensions. Below, we use the preceding procedure to find the joint PDF of the first three scheduled users' SINRs.
\subsubsection{Example}
\begin{figure}[!t]
\centering
\advance\leftskip-.15in
\vspace{-.075in}
\includegraphics[width=0.55\textwidth]{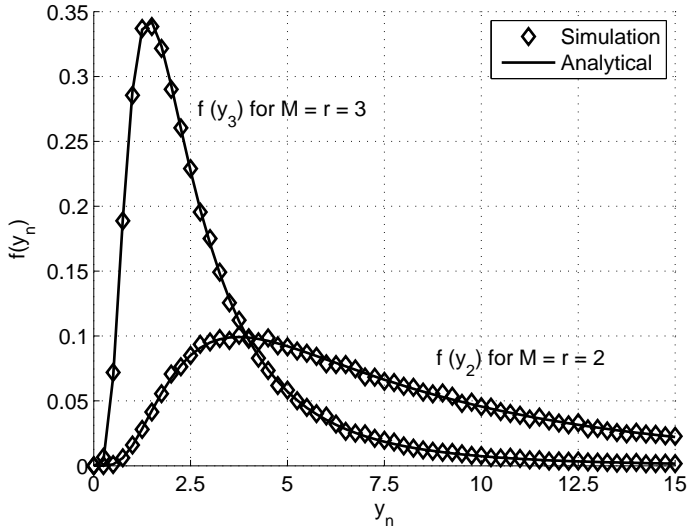}
\vspace{-3mm}
\caption{Comparison of the analytical PDF expressions with the numerical results for OLBF with $P=15$ dB and $K=10$.}
\label{OrtBF_Fig3:fig}
\end{figure}

We present an application of the general result given in Theorem~\ref{thm2:th} for the first three scheduled users' SINRs for any $K \geq M \geq 3$. We have $\xi_1(t_1)$ and $\xi_2(t_2,t_1)$ given by
\begin{displaymath}
\xi_1(t_1)=f(z_1=t_1)=\frac{e^{-\frac{M}{P}\frac{t_1}{1-t_1}}(M/P)^{M}}{(1-t_1)^{M+1}}
\frac{t_{1}^{M-1}}{(M-1)!}
\end{displaymath}
and
\begin{flalign}
\xi_2(t_2,t_1)=\int_{t_2}^{t_1}f(z_1,z_2=t_2) dz_1 && \nonumber
\end{flalign}
\begin{displaymath}
=\sum _{i=0}^{M-2} {M-2 \choose i}\frac{(-1)^i}{\Gamma(M-1)} \left(\frac{M}{P}
\right)^i e^{M/P} (1-t_{2})^{M-2-i}
\end{displaymath}
\vspace{-.1in}
\begin{flalign*}
&& \times \Bigg\{\Gamma\left(M-i,\frac{M}{P(1-t_{2})}
\right)-\Gamma\left(M-i,\frac{M}{P(1-t_{1})}\right)\Bigg\}.
\end{flalign*}
For $\xi_3(t_3,t_2,t_1)$, we have two upper bounds on $z_2$ as\linebreak $0 \leq z_2 \leq \{t_2, t_{1}-t_{3}\}$. Consequently, one can write
\begin{displaymath}
\xi_3(t_3,t_2,t_1)=\left\{ \begin{array}{lc}
\eta(t_2) & \mbox{ for  } t_1 \geq t_2 + t_3 ,\\
\eta(t_1 - t_3) & \mbox{ for  } t_1 < t_2 + t_3 ,
       \end{array} \right.
\end{displaymath}
where $\eta(.)$ is given in (\ref{eq:eta_x}) as a double-column equation. Using (\ref{eq:CDFz3}), $F_{z_{3}}(t_1,t_2,t_3)$ can be expressed as
\begin{flalign}
F_{z_{3}}(t_1,t_2,t_3) && \nonumber
\end{flalign}
\begin{equation}
=\left\{ \begin{array}{lc}
\mbox{Given in Appendix~\ref{appendixB}} & \mbox{ for  } t_1 \geq t_2 + t_3 ,\\
F_{z_{2}}(t_1,t_2)+F_{z_{2}}(t_1,t_3)-F_{z_{1}}(t_1) & \mbox{ for  } t_1 < t_2 + t_3 ,
       \end{array} \right.
       \label{eq:exampleCDFz3}
       \end{equation}
where
\begin{eqnarray}
F_{z_1}(t_1)&=&\int_{0}^{t_1}f(z_1)dz_1  \nonumber \\
&=& \int_{0}^{t_1} \frac{e^{-\frac{M}{P}\frac{z_{1}}{1-z_1}}(M/P)^{M}}{(1-z_1)^{M+1}}
\frac{z_{1}^{M-1}}{(M-1)!}~dz_1 \nonumber \\
&=&\sum_{i=0}^{M-1} {M-1 \choose i}\frac{(-1)^i}{\Gamma(M)} \left(\frac{M}{P}\right)^i e^{M/P}  \\
& & \times \left[\Gamma
\left(M-i,\frac{M}{P}\right)-\Gamma\left(M-i,\frac{M}{P(1-t_1)}\right)\right] \nonumber
\end{eqnarray}
and $F_{z_2}(t_1, t_2)$ is given in (\ref{eq:F_z2}) at the bottom of this page. Finally, the joint PDF of $\{t_1,t_2,t_3\}$ can be obtained by substituting $F_{z_{3}}(t_1,t_2,t_3)$, $\xi_1(t_1)$, $\xi_2(t_2,t_1)$, and $\xi(t_3,t_2,t_1)$ in (\ref{eq:teorem2}) with $n=3$. This is given by
\addtocounter{equation}{1}
\begin{equation}
f(t_{1},t_{2},t_{3})\!=\!\!\left\{ \begin{array}{l}
\!\!\!\frac{K!}{(K-3)!}\left(F_{z_{3},t_1 \geq t_2 + t_3}(t_1,t_2,t_3)
\right)^{K-3} \xi_{1}(t_1) \vspace{.025in} \\
\hspace{.55in} \times \xi_{2}(t_2,t_1)\eta(t_2) \mbox{~~for } t_1 \geq t_2 + t_3 , \vspace{.1in} \\
\!\!\!\frac{K!}{(K-3)!} \left(F_{z_{3},t_1 < t_2 + t_3}(t_1,t_2,t_3)
\right)^{K-3} \xi_{1}(t_1) \vspace{.025in}\\
\hspace{.3in} \times \xi_{2}(t_2,t_1)\eta(t_1-t_3) \mbox{~~for } t_1 < t_2 + t_3 ,
       \end{array} \right.
\end{equation}
\\
with $0 \leq \{t_2,t_3\} \leq t_1 \leq 1$ where $F_{z_{3},t_1 \geq t_2 + t_3}(t_1,t_2,t_3)$ and $F_{z_{3},t_1 < t_2 + t_3}(t_1,t_2,t_3)$ refer to two segments involved in the evaluation of $F_{z_{3}}(t_1,t_2,t_3)$ for $t_1 \geq t_2 + t_3$ and $t_1 < t_2 + t_3$, respectively. At this point, we can first obtain the marginal PDF of $t_n$ and then apply the inverse of the transformation given in (\ref{eq:vn_to_zn}) to get the marginal PDF of the actual SINR $y_n$. Other possibility is to obtain the joint PDF of actual SINRs through the same transformation:
\begin{equation}
f(y_{1},y_{2},y_{3})=\frac{f\left(t_{1}=\frac{y_{1}}{1+y_{1}},t_{2}=\frac{y_{2}}{1+y_{2}}
,t_{3}=\frac{y_{3}}{1+y_{3}
}\right)}{\left(1+y_{1}\right)^{2}\left(1+y_{2}\right)^{2}\left(1+y_{3}\right)^{2}}
\end{equation}
with $0 \leq \{y_{2},y_{3}\} \leq y_{1}$~\cite{Papoulis:02}. The corresponding average sum rate can be expressed as
\begin{displaymath}
\mathds{E}\left[\log(1+y_1)+\log(1+y_2)+\log(1+y_3)\right].
\end{displaymath}
The PDF expressions of the second and third scheduled users are illustrated in Fig.~\ref{OrtBF_Fig3:fig} together with the corresponding simulated histograms for $P=15$ dB and $K=10$. The simulated results are based on $10^5$ realizations and two $M$ values of $M=\{2,3\}$ are used with the number of scheduled users being equal to $M$. The strong match between analytical and simulated results validates the accuracy of the analysis above.
\vspace{1in}
\section{Numerical Results}
\label{numericalresults}
\begin{figure}[!t]
\centering
\advance\leftskip-.15in
\vspace{-.075in}
\includegraphics[width=0.55\textwidth]{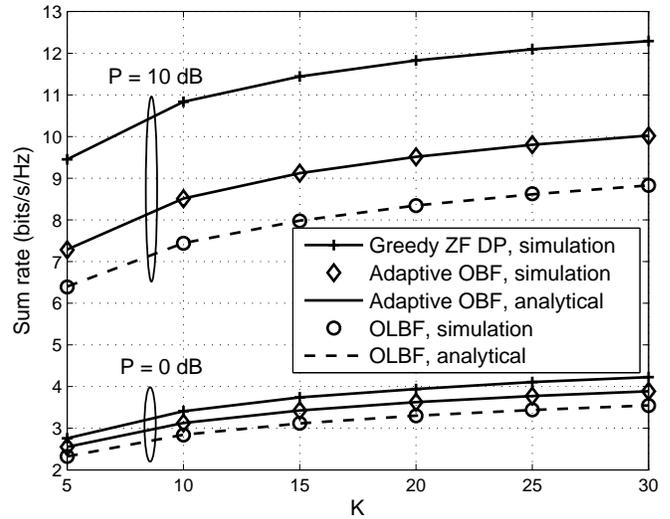}
\vspace{-3mm}
\caption{Sum rate performance of greedy ZF DP, adaptive OBF, and OLBF for $M=r=3$, $P=\{0, 10\}$ dB, and varying $K$.}
\label{OrtBF_Fig5:fig}
\vspace{-.1in}
\end{figure}

\begin{figure}[!t]
\centering
\advance\leftskip-.15in
\vspace{-.075in}
\includegraphics[width=0.55\textwidth]{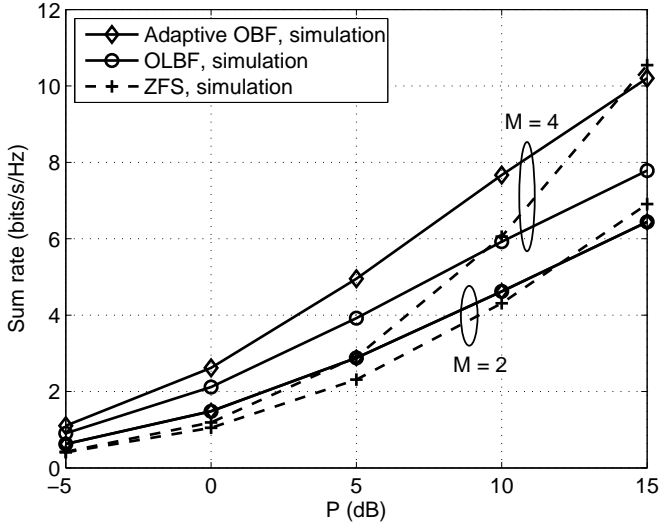}
\vspace{-3mm}
\caption{Sum rate comparison of adaptive OBF and OLBF with ZFS for $M=r=\{2, 4\}$, $K=M$, and varying $P$.}
\label{OrtBF_Fig4:fig}
\end{figure}

This section illustrates the sum rate performance of adaptive OBF and OLBF algorithms by computer simulations and analytical results. We also compare these two orthogonal BF schemes with two other well-known multiuser BF algorithms. The first one is greedy zero-forcing dirty paper coding (ZF DP) algorithm under uniform power allocation and is used as a benchmark scheme. Greedy ZF DP is a combination of scalar dirty paper coding and BF at the transmitter to yield interference-free users~\cite{Caire:03, Tu:03}. It has been shown to attain the same slope of sum rate increase with transmit power in dB as the capacity-achieving DPC~\cite{Dimic:05}. The other scheme used is based on zero-forcing BF and called zero-forcing BF with greedy user selection (ZFS)~\cite{Dimic:05}. In order to make a fair comparison, we assume that the number of scheduled users is $M$ for all four schemes. The average sum rate is depicted with respect to $K$ with $M=3$ and $P=\{0, 10\}$ dB in Fig.~\ref{OrtBF_Fig5:fig}. For $P=10$ dB, adaptive OBF and OLBF yield sum rates that are more than $75\%$ and $65\%$ of greedy ZF DP sum rate, respectively. The sum rate loss against greedy ZF DP is smaller for $P=0$ dB where the sum rates of adaptive OBF and OLBF are more than $90\%$ and $80\%$ of greedy ZF DP sum rate, respectively. For $P$ values around $0$ dB, one can use either of the low complexity algorithms adaptive OBF or OLBF without losing much in terms of sum rate performance. Most importantly, the simulated and analytical curves are on top of each other verifying the accuracy of the analytical derivations. In Fig.~\ref{OrtBF_Fig4:fig}, we plot the average sum rate for $M=\{2, 4\}$ and varying $P$ where the number of users is chosen as $K=M$. Note that adaptive OBF and OLBF become identical for $M=2$ under this setting. In this scenario, two orthogonal BF schemes outperform ZFS for most of the plotted $P$ range. The performance gap can be attributed to the better control of multiuser interference by orthogonal BF schemes~\cite{Francisco:07}. Also, adaptive OBF yields better performance as compared to OLBF under this setting.
\section{Conclusion}
\label{conclusion}
In this paper, we investigate a statistical analysis for the sum rate performance of two orthogonal BF schemes, each one of which carries out user scheduling and computation of BF weight vectors in conjunction. Multiuser BF algorithms combined with non-random user scheduling techniques are generally difficult to analyze as ordering among the user SINRs make it intractable to derive marginal PDFs of the scheduled users' SINRs by standard approaches using conditional PDFs. By taking a fresh look at the problem, we have derived a general result from order statistics on the joint PDF of scheduled users' SINRs under a certain scheduling algorithm with some definite properties. This result has been specifically applied to two orthogonal BF schemes, namely adaptive OBF and OLBF, to find compact solutions for the joint PDF of the scheduled users' SINRs. In particular, we have obtained closed-form expressions for the first three scheduled users for both schemes. Our key finding is to demonstrate the relation between the probability distribution functions in unordered and ordered SINR cases. Evaluation of the CDF of unordered SINRs has been shown to be instrumental for analysis and a recursive technique has been developed to evaluate the CDF in OLBF case. Finally, we have verified our analysis with numerical results. The presented analysis can also be applied for similar algorithms.
\vspace{.35in}
\appendices
\section{\texorpdfstring{Proof of Theorem~\ref{thm1:th}}{}}
\label{appendixA}
We resort to the result of Section~\ref{mainresult}. In order to evaluate the integral $\phi_k(y_k,y_{k-1},\ldots,y_1)$ in (\ref{eq:mainresult}), we need the region defined by $S_k$. For adaptive OBF, this is given by
\begin{equation}
S_k : v_{k1} \geq v_{k2} \geq \ldots \geq v_{kn} \geq 0
\label{eq:S_k}
\end{equation}
for all $k \in \{1,2,\ldots,K\}$. Using this together with (\ref{eq:sinirlar}), $\phi_k(y_k,y_{k-1},\ldots,y_1)$ can be written as
\begin{flalign}
\phi_k(y_k,y_{k-1},\ldots,y_1) && \nonumber
\end{flalign}
\begin{displaymath}
=\int_{y_k}^{y_{k-1}}\int_{v_{k-1}}^{y_{k-2}}\ldots\int_{v_2}^{y_1}f(v_1,v_2,\ldots,v_{k-1},v_k=y_k)
\end{displaymath}
\begin{equation}
\hspace{1.75in} \times dv_{1}\ldots dv_{k-2} dv_{k-1}
\label{eq:phi_k}
\end{equation}
for $k \in \{2, \ldots, n\}$. Note that due to the assumption that the network consists of homogeneous users, the PDF $f(v_{k1},v_{k2},\ldots,v_{kn})$ in (\ref{eq:compositejointpdf}) is independent of $k$. Also, $(K-n)$ CDF expressions inside the second parentheses in (\ref{eq:mainresult}) are identical. Consequently, the joint PDF of $\{y_1,y_2,\ldots,y_n\}$ can be written as
\begin{flalign}
f(y_1,\ldots,y_n)=\frac{K!}{(K-n)!}
\left(F_{v_{kn}}(y_1,y_2,\ldots,y_n)\right)^{K-n} && \nonumber
\end{flalign}
\begin{equation}
\hspace{1.75in} \times \prod_{k=1}^{n}\phi_k
(y_k,y_{k-1},\ldots,y_1).
\label{eq:esit5}
\end{equation}
Using (\ref{eq:sinirlar}) and (\ref{eq:S_k}) together with (\ref{eq:jointpdfofv}), the CDF expression in (\ref{eq:esit5}) can be evaluated as
\begin{flalign}
F_{v_{kn}}(y_1,y_2,\ldots,y_n) && \nonumber
\end{flalign}
\begin{eqnarray}
&=&\int_{0}^{y_n}\int_{v_n}^{y_{n-1}}\ldots\int_{v_2}^{y_1}f(v_1,v_2,\ldots,v_n)dv_{1}\ldots dv_{n-1} dv_{n}\nonumber\\
&=&\int_{0}^{y_n}\phi_{n}(\alpha,y_{n-1},\ldots,y_1)d\alpha
\label{eq:esit6}
\end{eqnarray}
where we apply a change of variables as $v_{n}=\alpha$. Using (\ref{eq:phi_k}) and (\ref{eq:esit6}) in (\ref{eq:esit5}) with a change of function names as $\phi\rightarrow\varphi$ yields the desired result given in (\ref{eq:teorem1}).
\section{\texorpdfstring{Evaluation of $F_{z_n}(t_1,t_2,\ldots,t_n)$ for $t_1 \! \geq \! t_2 \!+\! t_3 \!+\! \ldots \!+\! t_n$}{}}
\label{appendixB}
\vspace{.1in}
For $t_1 \geq t_2 + t_3 + \ldots + t_n$, $F_{z_{n}}(t_1,t_2,\ldots,t_n)$ can be written as
\begin{flalign}
F_{z_{n}}(t_1,t_2,\ldots,t_n) && \nonumber
\end{flalign}
\begin{flalign}
=\!\!\!\int\limits_{0}^{t_n}\!\ldots
\int\limits_{0}^{t_2}\int\limits_{z_2+z_3+\ldots+z_n}^{t_1} \!\!\!\!\!\!f(z_1,z_2,\ldots,z_n)~dz_1 dz_2 \ldots dz_n && \nonumber
\end{flalign}
\begin{flalign}
=\!\!\!\int\limits_{0}^{t_n}\!\ldots\int\limits_{0}^{t_2}\int\limits_{z_2+z_3+\ldots+z_n}^{t_1} \!\!\!\!\!\!\frac{e^{-\frac{M}{P}\frac{z_{1}}{1-z_{1}}}(M/P)^{M}}{(1-z_{1})^{M+1}}~\frac{\left(z_{1}
-\sum_{i=2}^{n}z_{i}\right)^{M-n}}{(M-n)!} && \nonumber
\end{flalign}
\begin{equation}
\hspace{2in} \times dz_1 dz_2 \ldots dz_n.
\end{equation}
\begin{figure*}[!b]
\addtocounter{equation}{3}
\hrule
\vspace{.15in}
\begin{eqnarray}
W_{n}(z_1,t_2,\ldots,t_n)&=&\left\{ \begin{array}{l}
\int_{0}^{t_n}\!\!\ldots\int_{0}^{t_2}f(z_1,z_2,\ldots,z_n)~ dz_2 \ldots dz_n\mbox{~~~~~~~for~~}  z_1 \geq t_2 + t_3+\ldots+t_n ,\vspace{.1in} \\
f(z_1)-\sum_{i=2}^{n}\overline{W}_{2}(z_1,t_i)\\
+\sum_{i=2}^{n-1}\sum_{j=i+1}^{n}\overline{W}_{3}(z_1,t_i,t_j)\\
\vdots\\
+(-1)^{n}\sum\limits_{\substack{2 \leq i_{1} < i_{2} <\\ \ldots <i_{n-2} \leq n}}\overline{W}_{n-1}(z_1,t_{i_1},\ldots,t_{i_{n-2}})
  \mbox{~~for~~} z_1 < t_2 + t_3+\ldots+t_n ,
       \end{array} \right.
       \label{eq:W_n} \\
       &~& \nonumber \\
       &~& \nonumber \\
\overline{W}_{n}(z_1,t_2,\ldots,t_n)&=&\left\{ \begin{array}{lc}
f(z_1)-\sum_{i=2}^{n}W_{2}(z_1,t_i)\\
+\sum_{i=2}^{n-1}\sum_{j=i+1}^{n}W_{3}(z_1,t_i,t_j)\\
\vdots\\
+(-1)^{n-1}W_{n}(z_1,t_{2},\ldots,t_{n})
 & \mbox{ for  } z_1 \geq t_2 + t_3+\ldots+t_n ,\vspace{.1in} \\
0 & \mbox{ for  } z_1 < t_2 + t_3+\ldots+t_n .
       \end{array} \right.
       \label{eq:W_n_head}
       \end{eqnarray}
       \vspace{.15in}
       \hrule
       \vspace{.15in}
       \begin{flalign}
\int_{0}^{t_n}\ldots\int_{0}^{t_3}\int_{0}^{t_2}f(z_1,z_2,\ldots,z_n) dz_2 \ldots dz_n && \nonumber
\end{flalign}
\begin{equation}
~~=\frac{e^{-\frac{M}{P}\frac{z_1}{1-z_1}}(M/P)^{M}}{(1-z_1)^{M+1}}~\frac{z_{1}^{M-1}-\sum\limits_{i=2}^{n}(z_1-t_{i})^{M-1}+\sum\limits_{2\leq i_{1} < i_{2} \leq n}(z_1-t_{i_{1}}-t_{i_{2}})^{M-1}+\ldots+(-1)^{n-1}\left(z_{1}-\sum\limits_{j=2}^{n}t_{j}\right)^{M-1}}{(M-1)!}.
\label{eq:esit7}
\end{equation}
\addtocounter{equation}{-6}
\end{figure*}
A direct closed-form solution for the above integral is rather complicated to derive. The difficulty occurs due to the fact that the inner-most integral is with respect to $z_1$. Instead of directly solving this integral, we change the integration order such that $z_1$ is the last integration variable and apply the inclusion-exclusion principle as similar to Section~\ref{OLBF}.\linebreak For $t_1 \geq t_2 + t_3 + \ldots + t_n$, we define $W_n(z_1,t_2,\ldots,t_n)$ and $\overline{W}_n(z_1,t_2,\ldots,t_n)$ as
\begin{displaymath}
W_n(z_1,t_2,\ldots,t_n)=\text{Pr~}(z_1, z_2 \leq t_2,z_3 \leq t_3, \ldots,z_n \leq t_n)
\end{displaymath}
and
\begin{displaymath}
\overline{W}_n(z_1,t_2,\ldots,t_n)=\text{Pr~}(z_1, z_2 > t_2,z_3 > t_3, \ldots,z_n > t_n).
\end{displaymath}
Then, we can write
\begin{equation}
F_{z_{n}}(t_1,t_2,\ldots,t_n)=\int_{z_1} W_n(z_1,t_2,\ldots,t_n) dz_1
\end{equation}
for $t_1 \geq t_2 + t_3 + \ldots + t_n$. Also, we have
\begin{displaymath}
\overline{W}_2(z_1,t_2)=\int_{t_2}^{z_1}f(z_1,z_2)dz_2
\end{displaymath}
and
\begin{displaymath}
W_2(z_1,t_2)=f(z_1)-\overline{W}_2(z_1,t_2)
\end{displaymath}
\linebreak
where $f(z_1)$ and $f(z_1, z_2)$ can be derived from (\ref{eq:jointpdfofzn}). For\linebreak $n=3$, we have
\begin{flalign}
W_3(z_1,t_2,t_3) && \nonumber
\end{flalign}
\begin{equation}
=\left\{ \begin{array}{lc}
\int_{0}^{t_3}\int_{0}^{t_2}f(z_1,z_2,z_3) dz_2 dz_3 & \mbox{\!\!for~~} z_1 \geq t_2 + t_3 ,\vspace{.1in} \\
f(z_1)-\overline{W}_2(z_1,t_2)-\overline{W}_2(z_1,t_3) & \mbox{\!\!for~~} z_1 < t_2 + t_3 ,
       \end{array} \right.
       \end{equation}
and
\newpage \noindent
\begin{flalign}
\overline{W}_{3}(z_1,t_2,t_3) && \nonumber
\end{flalign}
\begin{equation}
=\left\{ \begin{array}{l}
f(z_1)-W_2(z_1,t_2)-W_2(z_1,t_3)+W_3(z_1,t_2,t_3) \\
\hspace{2in}\mbox{~~for~~} z_1 \geq t_2 + t_3 ,\vspace{.1in} \\
0 \hspace{2in} \mbox{~for  } z_1 < t_2 + t_3.
       \end{array} \right.
      \end{equation}
~ \linebreak
Using the inclusion-exclusion principle [11], we can express $W_n(z_1,t_2,t_3,\ldots,t_n)$ and $\overline{W}_{n}(z_1,t_2,t_3,\ldots,t_n)$ in a recursive manner for any $n$ with $n \in \{3, \ldots,M \}$ as given in (\ref{eq:W_n}) and (\ref{eq:W_n_head}). The integral in the first line in (\ref{eq:W_n}) can be solved by substituting (\ref{eq:jointpdfofzn}). The closed-form solution is given in (\ref{eq:esit7}) which can be integrated over $z_1$ by applying a change of variables and using the binomial expansion formula~\cite{Gradshteyn:00}. The domain of the integration is given by $[ t_2 + t_3 + \ldots + t_n , t_1 ]$. The recursive structure presented above can be utilized to write $W_{n}(z_1,t_2,\ldots,t_n)$ with $n \in \{4,\ldots,M \}$ in terms of (\ref{eq:esit7}), $f(z_1)$, $W_{2}(z_1, t_i)$, and $W_{3}(z_1,t_{j},t_{k})$. Consequently, to evaluate $F_{z_n}(t_1,t_2,\ldots,t_n)$ for $t_1 \geq t_2 + t_3 + \ldots +t_n$, we only need to calculate the integrals of these simpler expressions over $z_1$ which can be evaluated by taking the segmentations of $z_1$ into account. As an example, the closed-form solution of $F_{z_3}(t_1,t_2,t_3)$ for $t_1 \geq t_2 + t_3$, i.e, $\int_{z_1} W_3(z_1,t_2,t_3) dz_1$, is given in (\ref{eq:F_z3}) at the top of the next page.
\addtocounter{equation}{3}
\begin{figure*}[!t]
\begin{flalign}
\int_{z_1} W_3(z_1,t_2,t_3) dz_1=\int_{t_2 + t_3}^{t_1}\int_{0}^{t_3}\int_{0}^{t_2}f(z_1,z_2,z_3) dz_2 dz_3 dz_1+
\int_{0}^{t_2 +t_3}f(z_1) dz_1 -\int_{t_2}^{t_2 +t_3}\int_{t_2}^{z_1}f(z_1,z_2)dz_2dz_1 \nonumber &&
\end{flalign}
\begin{displaymath}
 \hspace{3.35in} -\int_{t_3}^{t_2 +t_3}\int_{t_3}^{z_1}f(z_1,z_2)dz_2dz_1
\end{displaymath}
\begin{flalign}
\hspace{1.25in}=\sum_{i=0}^{M-1}{M-1 \choose i}\frac{(-1)^i}{\Gamma(M)}\left(\frac{M}{P}\right)^{i} e^{M/P}\Bigg\{\Gamma\left(M-i,\frac{M}{P}\right)-(1-t_2)^{M-i-1}\Gamma\left(M-i
,\frac{M}{P(1-t_2)}\right)  \nonumber &&
\end{flalign}
\begin{displaymath}
\hspace{1in} -(1-t_3)^{M-i-1}\Gamma\left(M-i
,\frac{M}{P(1-t_3)}\right)+(1-t_2-t_3)^{M-i-1}\Gamma\left(M-i
,\frac{M}{P(1-t_2-t_3)}\right)
\end{displaymath}
\begin{equation}
\hspace{1.2in} -\Bigg[1-(1-t_2)^{M-i-1}
-(1-t_3)^{M-i-1}+(1-t_2-t_3)^{M-i-1}\Bigg]\Gamma\left(M-i
,\frac{M}{P(1-t_1)}\right)\Bigg\}.
\label{eq:F_z3}
\end{equation}
\vspace{.25in}
\hrule
\end{figure*}
\hspace{-.1in} In the same manner, using (\ref{eq:W_n}), (\ref{eq:W_n_head}), and (\ref{eq:esit7}), $F_{z_n}(t_1,t_2,\ldots,t_n)$ for $t_1 \geq t_2 + t_3 + \ldots +t_n$ can be evaluated for any $n$ in an exact closed-form.

\begin{biography}[{\includegraphics[width=1in,height=1.25in,clip,keepaspectratio]{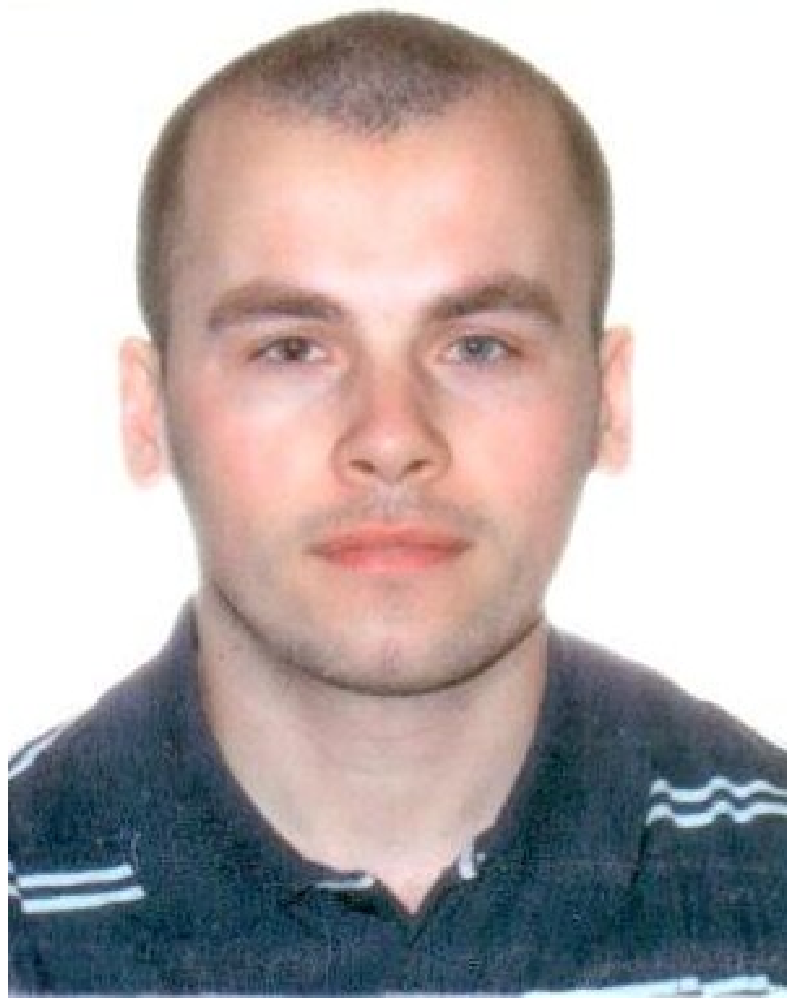}}]{Serdar Ozyurt}
received the B.Sc. degree in electrical
and electronics engineering from Sakarya University, Turkey, in 2001, the M.Sc. degree in
electronics engineering from Gebze Institute of Technology, Turkey, in 2005, and the
Ph.D. degree in electrical engineering from University of Texas at Dallas, USA, in 2012. He is now an Assistant Professor in the Department of Energy Systems Engineering, Yildirim Beyazit University, Ankara, Turkey. His current research interests include multiantenna communication systems with multiple users and related signal processing techniques.
\end{biography}
\begin{biography}[{\includegraphics[width=1in,height=1.25in,clip]{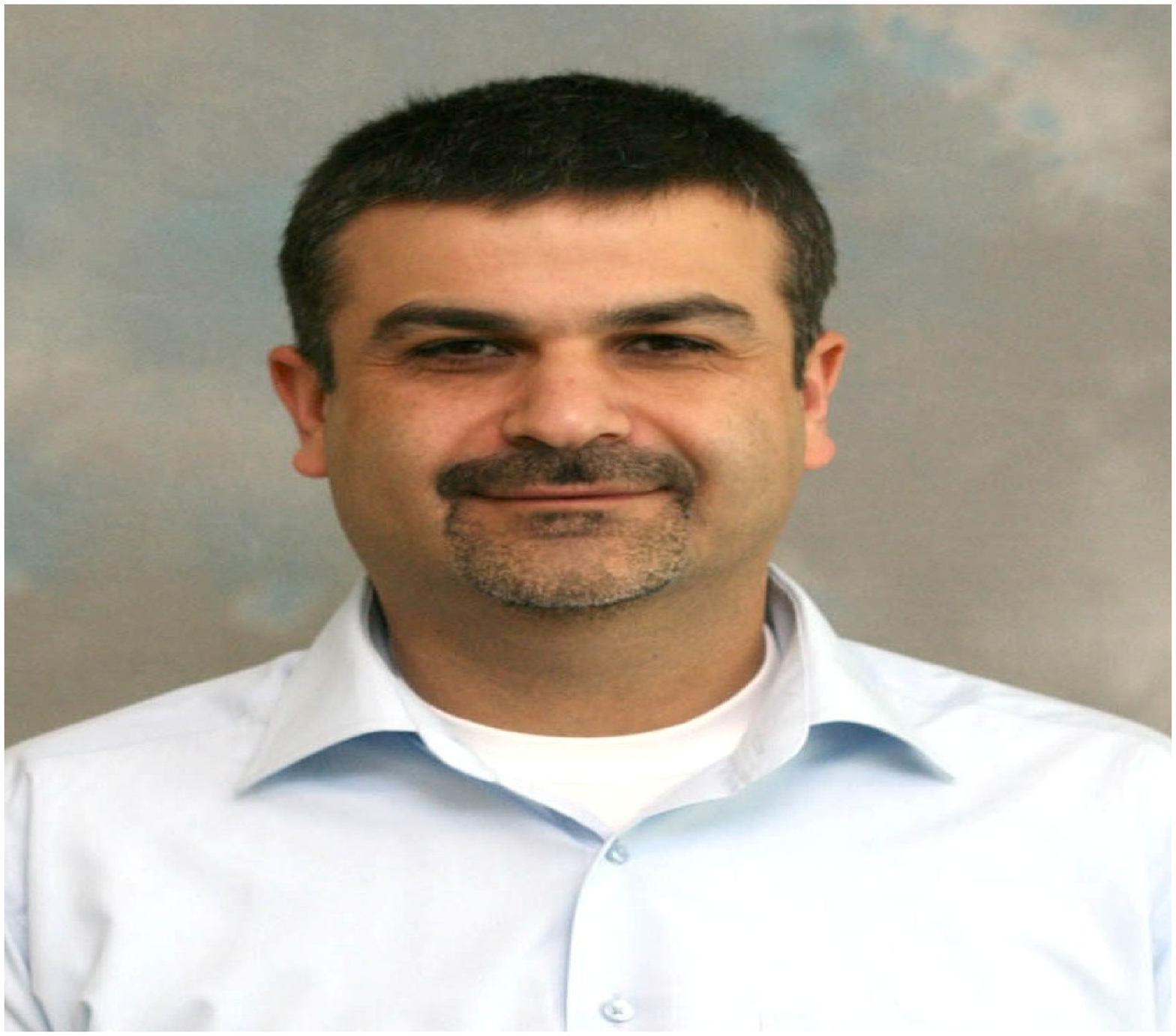}}]{Murat Torlak}
received M.S. and Ph.D. degrees in electrical engineering from The University of Texas at Austin in 1995 and 1999, respectively.  He is an Associate Professor in the Department of Electrical Engineering at the Eric Johnson School of Engineering and Computer Science within the University of Texas at Dallas.  He spent the summers of 1997 and 1998 in Cwill Telecommunications, Inc., Austin, TX, where he participated in the design of a smart antenna wireless local loop system and directed research and development efforts towards standardization of TD-SCDMA for the International Telecommunication Union. He was a visiting scholar at University of California Berkeley during 2008. He has been an active contributor in the areas of smart antennas and multiuser detection. His current research focus is on experimental platforms for multiple antenna systems, millimeter wave systems, and wireless communications with health care applications. He is an Associate Editor of IEEE Trans. on Wireless Communications. He was the Program Chair of IEEE Signal Processing Society Dallas Chapter during 2003-2005. He is a Senior IEEE member. He has served on the Technical Program Committees (TPC) of the several IEEE conferences.
\end{biography}
\end{document}

%% file: save2.pstex_t
\begin{picture}(0,0)%
\includegraphics{save2.pstex}%
\end{picture}%
\setlength{\unitlength}{3947sp}%
\begingroup\makeatletter\ifx\SetFigFont\undefined%
\gdef\SetFigFont#1#2#3#4#5{%
  \reset@font\fontsize{#1}{#2pt}%
  \fontfamily{#3}\fontseries{#4}\fontshape{#5}%
  \selectfont}%
\fi\endgroup%
\begin{picture}(6787,4387)(136,-4052)
\put(4501,-2461){\makebox(0,0)[lb]{\smash{{\SetFigFont{12}{14.4}{\rmdefault}{\mddefault}{\updefault}{\color[rgb]{0,0,0}$t_3$}%
}}}}
\put(2032,-2137){\makebox(0,0)[lb]{\smash{{\SetFigFont{12}{14.4}{\rmdefault}{\mddefault}{\updefault}{\color[rgb]{0,0,0}$z_2 + z_3 = t_1$}%
}}}}
\put(3024,-3176){\makebox(0,0)[lb]{\smash{{\SetFigFont{12}{14.4}{\rmdefault}{\mddefault}{\updefault}{\color[rgb]{0,0,0}$z_2$}%
}}}}
\put(5953,-2657){\makebox(0,0)[lb]{\smash{{\SetFigFont{12}{14.4}{\rmdefault}{\mddefault}{\updefault}{\color[rgb]{0,0,0}$z_2 + z_3 = t_1$}%
}}}}
\put(4585,-1570){\makebox(0,0)[lb]{\smash{{\SetFigFont{12}{14.4}{\rmdefault}{\mddefault}{\updefault}{\color[rgb]{0,0,0}$z_3$}%
}}}}
\put(6662,-3223){\makebox(0,0)[lb]{\smash{{\SetFigFont{12}{14.4}{\rmdefault}{\mddefault}{\updefault}{\color[rgb]{0,0,0}$z_2$}%
}}}}
\put(993,-1617){\makebox(0,0)[lb]{\smash{{\SetFigFont{12}{14.4}{\rmdefault}{\mddefault}{\updefault}{\color[rgb]{0,0,0}$z_3$}%
}}}}
\put(4050,-4050){\makebox(0,0)[lb]{\smash{{\SetFigFont{12}{14.4}{\rmdefault}{\mddefault}{\updefault}{\color[rgb]{0,0,0}$\overbrace{F_{z_1}(t_1)-\overline{F}_{z_2}(t_1,t_2)-\overline{F}_{z_2}(t_1,t_3)}$}%
}}}}
\put(180,-4050){\makebox(0,0)[lb]{\smash{{\SetFigFont{12}{14.4}{\rmdefault}{\mddefault}{\updefault}{\color[rgb]{0,0,0}$\overbrace{\displaystyle{\int_{0}^{t_3}\int_{0}^{t_2}\int_{z_2 + z_3}^{t_1}f(z_1,z_2,z_3)dz_1 dz_2 dz_3}}$}%
}}}}
\put(3226,164){\makebox(0,0)[lb]{\smash{{\SetFigFont{16}{16.8}{\rmdefault}{\mddefault}{\updefault}{\color[rgb]{0,0,0}$F_{z_3}(t_1,t_2,t_3)$}%
}}}}
\put(1726,-3361){\makebox(0,0)[lb]{\smash{{\SetFigFont{12}{14.4}{\rmdefault}{\mddefault}{\updefault}{\color[rgb]{0,0,0}$t_2$}%
}}}}
\put(5626,-3361){\makebox(0,0)[lb]{\smash{{\SetFigFont{12}{14.4}{\rmdefault}{\mddefault}{\updefault}{\color[rgb]{0,0,0}$t_2$}%
}}}}
\put(5451,-961){\makebox(0,0)[lb]{\smash{{\SetFigFont{14}{16.8}{\rmdefault}{\mddefault}{\updefault}{\color[rgb]{0,0,0}$t_1 < t_2 + t_3$}%
}}}}
\put(1551,-961){\makebox(0,0)[lb]{\smash{{\SetFigFont{14}{16.8}{\rmdefault}{\mddefault}{\updefault}{\color[rgb]{0,0,0}$t_1 \geq t_2 + t_3$}%
}}}}
\put(901,-2536){\makebox(0,0)[lb]{\smash{{\SetFigFont{12}{14.4}{\rmdefault}{\mddefault}{\updefault}{\color[rgb]{0,0,0}$t_3$}%
}}}}
\end{picture}%

%% file: TW-Dec-11-2304.bbl
\newpage \noindent
\begin{thebibliography}{10}
\bibitem{Weingarten:06}
H.~Weingarten, Y.~Steinberg, and S.~Shamai, ``The capacity region of the Gaussian multiple-input multiple-output broadcast channel," \emph{IEEE Trans. Inf. Theory}, vol. 52, no. 9, pp. 3936-3964, Sept. 2006.

\bibitem{Yin:02}
H.~Yin and H.~Liu, ``Performance of space-division multiple-access (SDMA) with scheduling," \emph{IEEE Trans. Wireless Commun.}, vol. 1, no. 4, pp. 611-618, Oct. 2002.

\bibitem{Yoo:06}
T.~Yoo and A.~Goldsmith, ``On the optimality of multiantenna broadcast scheduling using zero-forcing beamforming," \emph{IEEE J. Sel. Areas Commun.}, vol. 24, no. 3, pp. 528-541, Mar. 2006.

\bibitem{Dimic:05}
G.~Dimic and N.~D. Sidiropoulos, ``On downlink beamforming with greedy user selection: performance analysis and a simple new algorithm," \emph{IEEE Trans. Signal Process.}, vol. 53, no. 10, pp. 3857-3868, Oct. 2005.

\bibitem{Francisco:07}
R.~de Francisco, M.~Kountouris, D.~T. M. Slock, and D.~Gesbert, ``Orthogonal linear beamforming in MIMO broadcast channels," in \emph{Proc. IEEE Wireless Commun. Netw. Conf.,} Hong Kong, China, Mar. 2007, pp. 1210-1215.

\bibitem{Duplicy:07}
J.~Duplicy, D.~P. Palomar, and L.~Vandendorpe, ``Adaptive orthogonal beamforming for the MIMO broadcast channel," in \emph{Proc. IEEE Int. Work. Computational Adv. Multi-Sensor Adaptive Processing,} Virgin Islands, USA, Dec. 2007, pp. 77-80.

\bibitem{Ozdemir:10}
O.~Ozdemir and M.~Torlak, ``Optimum feedback quantization in an opportunistic beamforming scheme," \emph{IEEE Trans. Wireless Commun.}, vol. 9, no. 5, pp. 1584-1593, May 2010.

\bibitem{Ozyurt&VBLAST:11}
S.~Ozyurt and M.~Torlak, ``An exact outage analysis of zero-forcing\linebreak V-BLAST with greedy ordering," in \emph{Proc. IEEE Global Commun. Conf.-Broad. Wireless Access Work.,} Houston, USA, Dec. 2011.

\bibitem{Ozyurt:12}
S.~Ozyurt and M.~Torlak, ``Performance analysis of optimum zero-forcing beamforming with greedy user selection," \emph{IEEE Commun. Lett.}, vol. 16, no. 4, pp. 446-449, Apr. 2012.

\bibitem{Huang:09}
K.~Huang, J.~G. Andrews, and R.~W. Heath Jr., ``Performance of orthogonal beamforming for SDMA with limited feedback," \emph{IEEE Trans. Veh. Technol.}, vol. 58, no. 1, pp. 152-164, Jan. 2009.

\bibitem{Yang:11}
H.~Yang, P.~Lu, H.~Sung, and Y.~Ko, ``Exact sum-rate analysis of MIMO broadcast channels with random unitary beamforming,"  \emph{IEEE Trans. Commun.}, vol. PP, no. 99, pp. 1-5, June 2011.

\bibitem{David:03}
H.~A. David and H.~N. Nagaraja, \emph{Order Statistics}, 3rd ed. NJ: Wiley, 2003.

\bibitem{Sharma:05}
N.~Sharma and L.~Ozarow, ``A study of opportunism for multiple-antenna systems," \emph{IEEE Trans. Inf. Theory}, vol. 51, no. 5, pp. 1804-1814, May 2005.

\bibitem{Gradshteyn:00}
I.~Gradshteyn and I.~Ryzhik, \emph{Table of Integrals, Series, and Products}, San Diego: Academic Press, 2000.

\bibitem{Papoulis:02}
A.~Papoulis and S.~U. Pillai, \emph{Probability, Random Variables and Stochastic Processes}, 4th ed. NY: McGraw Hill, 2002.

\bibitem{Ozyurt:11}
S.~Ozyurt and M.~Torlak, ``Performance analysis of orthogonal beamforming with user selection in MIMO broadcast channels," in \emph{Proc. IEEE Global Commun. Conf.,} Houston, USA, Dec. 2011.

\bibitem{Ng:11}
K.~W. Ng, G.-L. Tian, and M.-L. Tang, \emph{Dirichlet and Related Distributions: Theory, Methods and Applications}, NY: Wiley Series in Probability and Statistics, 2011.


\bibitem{Gouda:10}
A.~A. Gouda and T.~Szantai, ``On numerical calculation of probabilities according to Dirichlet distribution," \emph{Annals Operations Research}, vol. 177, no. 1, pp. 185-200, June 2010.

\bibitem{Szantai:85}
T.~Szantai, ``Numerical evaluation of probabilities concerning multidimensional probability distributions," Thesis, \emph{Hungarian Academy of Sciences}, Budapest, Hungary, 1985.

\bibitem{Caire:03}
G. Caire and S. Shamai, ``On the achievable throughput of a multi-antenna Gaussian broadcast channel," \emph{IEEE Trans. Inf. Theory}, vol. 49, no. 7, pp. 1691–1706, July 2003.

\bibitem{Tu:03}
Z.~Tu and R. S. ~Blum, ``Multiuser diversity for a dirty paper approach," \emph{IEEE Commun. Lett.}, vol. 7, no. 8, pp. 370-372, Aug. 2003.
\end{thebibliography}
